\documentclass{aa}  
\usepackage[pdfpagelabels=false]{hyperref}	
\hypersetup{colorlinks=true,linkcolor=blue,citecolor=blue,filecolor=blue,urlcolor=blue,}
\usepackage{graphicx}
\usepackage{txfonts}
\usepackage{lipsum}
\usepackage{subcaption}         
\usepackage{lscape}             
\usepackage{placeins}           

 \usepackage{mhchem}
\usepackage{txfonts}

\begin{document}

   \title{The properties of central stellar knots embedded in galactic bulges of CIELO simulations}
   \subtitle{}

%

   \author{Belén Acosta-Tripailao\inst{1}\fnmsep\thanks{Corresponding author: \href{mailto:bacostr@uc.cl}{bacostr@uc.cl}} 
        \and Patricia B. Tissera\inst{1}\fnmsep\inst{2}
        \and Manuela Zoccali\inst{1}
        \and Jenny Gonzalez-Jara\inst{1}\fnmsep\inst{2}
        \and Brian Tapia-Contreras \inst{1}\fnmsep\inst{2}
        \and Ignacio Muñoz-Escobar\inst{1}\fnmsep\inst{2}
        \and Susana Pedrosa\inst{3}
        \and Nelson D. Padilla\inst{4}
        \and Lucas Bignone\inst{3}
        \and Rosa Dominguez-Tenreiro\inst{5}
        }

   \institute{Instituto de Astrofísica, Pontificia Universidad Católica de Chile, Av. Vicuña Mackenna, 4860 Santiago, Chile.
    \and Centro de AstroIngeniería, Pontificia Universidad Católica de Chile. Av. Vicuña Mackenna 4860, Santiago, Chile.    
    \and Instituto de Astronom\'ia y F\'isica del Espacio, CONICET-UBA, Casilla de Correos 67, Suc. 28, 1428, Buenos Aires, Argentina.
    \and Instituto de Astronom\'ia Te\'orica y Experimental, Laprida 900, C\'ordoba, Argentina.
    \and Departamento de Física Teórica, Universidad Autónoma de Madrid, E-28049 Cantoblanco, Madrid, Spain.
    }


   \date{Received May 18, 2026}

  \abstract
   {Deeply bound stellar substructures of about 1 kpc at the center of bulges are found in a subset of galaxies from the cosmological chemo-hydrodynamical zoom-in \textsc{CIELO} simulation suite. They were identified as stellar overdensities at the lowest binding energies in each galaxy's circularity-energy $(\epsilon, E)$ plane. We refer to these overdensities as stellar knots.}
   {We aim to characterize the nature and origin of the central stellar knots in the context of bulge assembly using \textsc{CIELO} simulated galaxies spanning a wide range of stellar masses ($10^{8.0}$--$10^{10.7}\,\mathrm{M}_{\odot}$), with diverse formation histories.} 
   {We inspect 54 galactic bulges. Within them, we isolate stellar knot candidates in $(\epsilon, E)$ space, and identify 28 robust knots satisfying successive selection criteria: kinematic, concentration, and morphology. We characterize their chemical enrichment, formation timescales, formation sites, progenitor gas origin, and spatial distributions.}
    {We detect some systematic differences in the median statistics of the CIELO knots relative to their local and global surrounding bulge stellar populations, although substantial scatter exists within the sample. Across all galaxy masses, knots are systematically alpha-element enhanced, having assembled the bulk of their stellar mass at earlier epochs and on shorter timescales than other bulge populations, with a median of $\sim$2$\,$Gyr versus $\sim$5$\,$Gyr. Regarding their origin, knots are predominantly in-situ ($5\%$ accreted mass fraction, roughly half that of the rest of the bulge), with negligible disk-born stars contribution and the largest gas fractions originating from a primordial central spheroid ($21\%$ versus 16--17$\%$ for the surrounding populations), consistent with being the primary gas fuel for the alpha-element enhancement. In terms of structure, knots exhibit a variety of morphologies, with spheroidal shapes predominating.}
   {Dynamical selection in $(\epsilon, E)$ space of CIELO galaxies demonstrates its effectiveness in recovering coeval stellar populations, pointing to stellar knots as plausible fossil signatures of early in-situ bulge assembly.}

    \titlerunning{The properties of central stellar knots}
    \maketitle
    \nolinenumbers
\section{Introduction}
In a $\Lambda$ cold dark matter ($\Lambda$CDM) universe, the growth of the dark matter halos is driven by gravity in a hierarchical fashion~\citep{davis+1985}. The resulting gravitational potential wells provide the sites where baryons cool and condense, leading to star formation. Throughout galaxy evolution, energy and chemical feedback processes are sensitive to the depth of the galaxy's gravitational potential well~\citep{somerville_dave2015, scannapieco+2005, scannapieco+2006}, with the efficiency of Supernova (SN) and Active Galactic Nucleus (AGN) feedback being strongly mass-dependent. For instance, the shallowness of the galactic potential well in low mass systems allows SN and AGN feedback energy to exceed the binding energy of gas, determining the strength of galactic winds or outflows and thus affecting the efficiency of baryon retention~\citep{dekel+silk1986, white_frenk1991}. Galaxy mass is one of the main drivers of galaxy evolution as suggested by the Mass-Size relation \citep{vanderwel2014}, the Star Formation Rate-Mass relation (Star-Forming Main Sequence; \citealt{speagle2014,baker2023}), the Mass-Metallicity relation \citep{tremonti2004,Gallazzi2005}, among others. This connection between mass and galaxy evolution even extends to the central supermassive black hole, whose mass correlates with several bulge and galaxy properties~(\citealp{ferrarese+merrit2000,gultekin+2009}; \citealp{kormendy+ho2013}, and references therein), despite the vastly different physical scales involved. As galaxies formed within their dark matter halos, their central regions became more bound and concentrated~\citep{white_rees1978}. Hence, understanding how mass is distributed in central regions of galaxies and how it shapes the kinematics of stellar populations is crucial for galactic archaeology.

Stellar populations in the central region of a galaxy are expected to encode key information about its early evolutionary stages. Theoretical and observational evidence suggests that this inner region assembles first, retaining signatures of early formation processes~\citep[e.g.][]{white_springel2000, scannapieco+2006, starkenburg+2017}. These signatures are encoded in the internal chemical and dynamical features of bulge stellar populations, such as the alpha-element enhancement, morphology, kinematics, among others~\citep{aguerri+2001,tissera+2011, martig+2021}. Given that our Galaxy is accessible to detailed study, the Milky Way (MW) Bulge provides the clearest observational evidence suggesting that the central region is the first massive component to assemble~(\citealp{renzini2018, savino2020, hasselquist2020, joyce2023}; \citealp{zoccali+valenti2024}, and references therein).

The present work is motivated by the reported central peak in the line-of-sight velocity dispersion ($\sigma$-peak) within the MW Bulge~\citep{zoccali+2014, valenti+2018, quezada2025}, reaching ${\sim}140\,\mathrm{km\,s^{-1}}$ within the inner few degrees ($\sim$280$\,$pc). This peak appears to be driven by metal-rich stars~\citep[$|b| \lesssim 5^{\circ}$;][]{babusiaux+2014, babusiaux2016, zoccali+2017}, though more precise data must confirm this. Possible explanations include a strong central mass concentration, as velocity dispersion may trace mass by the virial theorem, or orbit anisotropy from elongated orbits aligned with the LOS due to the bar's orientation. A spatially concentrated, round, non-rotating, and predominantly metal-rich population has also been found and referred to as a tight ``knot'' of stars~\citep{rix+2024_mrichknot, horta+2025_knot}. While chemically selected, this population appears centrally clustered~\citep{rix+2024_mrichknot}, but shows no clear chemical differences from its surroundings when defined dynamically~\citep{horta+2025_knot}, raising the question of whether it constitutes a genuinely distinct structure. If confirmed as such, understanding its origin could shed light on bulge assembly processes that may also operate in other galaxies, not only in the MW.

High-resolution zoom-in simulations of MW mass-sized halos offer an ideal framework for studying the assembly of galactic bulges. \citet{tissera+zoccali+2017} analyzed the central spheroid region within the inner $10$ kpc of the Aquarius sample, finding kinematics for accreted and in-situ stars consistent with \citet{zoccali+2017}. The Auriga project~\citep{grand+2017_auriga} reproduces MW-like stellar masses, metallicities, and rotation curves, with some analogues hosting bars~\citep{fragkoudi+2025} and boxy/peanut bulges~\citep{lopez+2025}. Interestingly, even though the spheroidal component makes up about half of the MW bulge's stellar mass~\citep{zoccali+2018_weighing}, simulations rarely produce spheroids with significant ex-situ contribution~\citep{gargiulo+2019, fragkoudi+2020_mwquiescent}, likely reflecting quiescent formation histories with the last major merger $\sim$12$\,$Gyr ago. Recently, \citet{munoz-escobar2026} studied 44 \textsc{CIELO} bulges~\citep{tissera+2025}, finding in-situ formation dominant, though a third of bulges show a non-negligible accreted contribution (from $20\%$ up to $35\%$), and disk migration contributes a median of $10\%$, up to $32\%$ in the most massive bulges.

Beyond these global properties, the internal structure of \textsc{CIELO} bulges reveals an additional feature: an apparent overdensity in the distribution of stellar particles at the low-energy end of the circularity-energy $(\epsilon, E)$ plane tracing a deeply bound population embedded at the galactic center, as shown in Fig.~6 from \citet{tissera+2025}. We refer to these stellar overdensities as knots. Sitting in the deepest regions of the gravitational potential well, knots are expected to reflect local mass concentrations. If such structures correspond to stellar populations sharing a common evolutionary history, they could provide valuable insight into the assembly of the innermost regions of galaxies. Clustering techniques can uncover such underlying chemo-dynamical diversity when applied with appropriate care~\citep{deleo+2025_dynamicalmirages, dillamore2025}, and groups of stars with similar chemo-dynamical properties are commonly interpreted as having formed from the same gas reservoir~\citep{jofre2017_phylogeny, jackson2021, gudin2021, tapia-contreras+2026}. The $(\epsilon, E)$ plane thus offers a natural starting point for uncovering central stellar knots and characterizing their chemical enrichment, with the aim of constraining their origin.

In this work, we aim to understand the nature and origin of central stellar knots using state-of-the-art hydrodynamical zoom-in simulations from the \textsc{CIELO} project, approaching this as an exploratory study of plausible trends in these substructures. For this purpose, we decided to use the \textsc{CIELO} zoom-ins suite. Cosmological hydrodynamical simulations describe the formation and evolution of galaxies in a coherent way within a cosmological context. \textsc{CIELO} offers a detailed chemical evolution model, tracking individual elements and isotopes from SNIa and SNII with empirically validated delay-time distributions~\citep{jimenez+2015, tissera+2025}, which enables a robust characterization of the chemical signatures of stellar populations. These simulations have been used to study the bulge mass-metallicity relation~\citep{munoz-escobar2026}, the metallicity distribution in the star-forming gas~\citep{tapia-contreras+2025}, the chemical abundances of stellar haloes and their connection to the assembly histories~\citep{gonzalez-jara+2025}, among others.

Through a distinct selection approach based on the $(\epsilon, E)$ plane, we provide evidence for a general class of centrally concentrated stellar overdensities in simulated galaxies, rather than a direct numerical counterpart to the knot reported in the MW Bulge. In a companion paper (Acosta-Tripailao et al., in prep.), we will move into the observational space to investigate the role of knots in the $\sigma$-peak of bulges and their interplay with the presence of galactic bars. The \textsc{CIELO} sample mainly includes galaxies with stellar mass ranging from $10^{8.0}$ to $10^{10.7}\,\mathrm{M}_{\odot}$, with the three most massive galaxies falling within the MW stellar mass range~\citep{lian_2025, pillepich_2024}. Hence, our purpose is to investigate the existence of knots and their characteristics for galaxies with a variety of masses and formation histories.

This paper is organized as follows. Section~\ref{sec:CIELO} describes the \textsc{CIELO} simulations, and Section~\ref{sec:knotsincielo} details our selection criteria for identifying central mass concentrations. In Section~\ref{sec:charact} we investigate their chemical abundance signatures, formation timescales, morphology, and the relative contributions of secular versus ex-situ processes to their assembly. We compare our results with other works in Section~\ref{sec:discu} and present our conclusions in Section~\ref{sec:conc}.

\section{CIELO simulations}
\label{sec:CIELO}
\begin{figure}
    \centering
    \includegraphics[width=\linewidth]{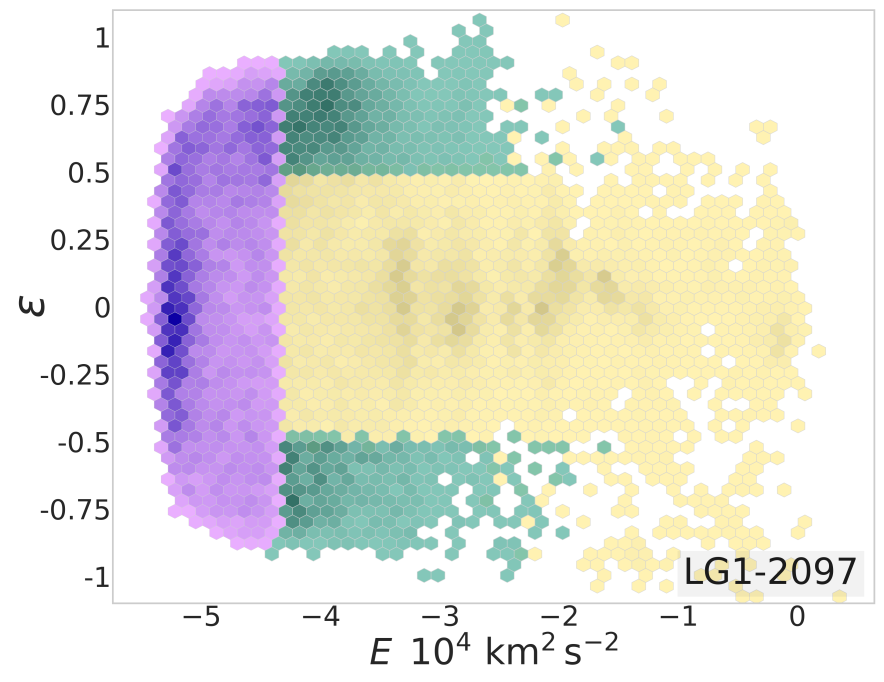}
    \caption{Circularity--energy plane of stellar particles from galaxy LG1-2097, classified as bulge, disk, and halo via the AM--E method~\citep{tissera2012}, shown as hexbins. Bulge particles range from purple (low density) to dark blue (high density). A prominent overdensity is visible at the lowest energies, the primary focus of this work. Disk and halo particles are shown in green and yellow, respectively, with darker shades indicating higher density.}
    \label{fig:am-e}
\end{figure}

The \textsc{CIELO} project~\citep{tissera+2025} investigates the formation and chemical evolution of galaxies across different cosmic environments, based on cosmological hydrodynamical zoom-in simulations consistent with a $\Lambda$CDM cosmology ($\Omega_0 = 0.3170$, $\Omega_\Lambda = 0.6825$, $\Omega_B = 0.049$, $h = 0.6711$). Simulated regions were selected from dark matter-only parent simulations: a 100~Mpc$\,$h$^{-1}$ box for the Local Group analogues (LG1 and LG2), and a 50~Mpc$\,$h$^{-1}$ box for the Pehuen halos (P), spanning virial masses of $10^{10}$--$10^{12}\,\mathrm{M}_\odot$, with initial conditions generated using the \textsc{MUSIC} code~\citep{hahn&abel2011}.

The suite comprises two resolution levels. L11 runs have dark matter and gas particle masses of $\rm m_{\rm dm}=1.36\times10^{6}$ and $\rm m_{\rm gas}=2.0\times10^{5}\,\mathrm{M}_\odot\,h^{-1}$, respectively, while L12 adopts $\rm m_{\rm dm} = 1.36\times10^{5}$ and $\rm m_{\rm gas}=2.1\times10^{4}\,\mathrm{M}_\odot\,h^{-1}$. Gravitational softening lengths are $\epsilon_\mathrm{grav} = 800$ and $500\,$pc for dark matter, and $400$ and $250\,$pc for gas and stellar particles, for L11 and L12, respectively. The suite contains 128 snapshots from $z\sim60$ to $z=0$, with a time cadence of $\leq$160 Myr.

The simulations were performed with a modified {\small GADGET-3}, based on {\small GADGET-2}~\citep{springel2003, springel2005}, incorporating a multiphase interstellar medium, metal-dependent radiative cooling, stochastic star formation, and SNIa/SNII feedback, following \citet{scannapieco+2005, scannapieco+2006}. Chemical enrichment tracks twelve isotopes: H, \ce{^{4}He}, \ce{^{12}C}, \ce{^{14}N}, \ce{^{16}O}, \ce{^{20}Ne}, \ce{^{24}Mg}, \ce{^{28}Si}, \ce{^{32}S}, \ce{^{40}Ca}, \ce{^{56}Fe}, and \ce{^{62}Zn}~\citep{mosconi+2001}. For SNIa, the W7 model~\citep{iwamoto+1999} is adopted, assuming progenitor lifetimes randomly distributed in the range 0.7--1.1 Gyr (see \citealt{jimenez+2015} for a detailed discussion). Gas initially has primordial abundances ($X_{\rm H} = 0.76$, $Y_{\rm He} = 0.24$, $Z = 0$), and the multiphase ISM/feedback scheme self-consistently drives strong, mass-loaded galactic winds without explicit mass-scale parameters~\citep{scannapieco2008}. A Chabrier IMF~\citep{chabrier2003} is adopted; virial haloes were identified via Friends-of-Friends~\citep{davis+1985}, subhaloes with \textsc{SUBFIND}~\citep{springel+2001, dolag+2009}, and merger trees with \textsc{AMIGA}~\citep{Knollmann_2009}.

\citet{tissera+2025} applied a dynamical decomposition of stellar particles, separating the halo, the disc, and the bulge over \textsc{CIELO} galaxies. The method, so-called AM-E~\citep{tissera2012}, is based on the total binding energy $E$ and the angular momentum content through the circularity parameter $\epsilon = J_z/J_{z,\mathrm{max}}(E)$, where $J_{z,\mathrm{max}}$ is the maximum angular momentum among particles with a given total energy $E$ (see Fig.~\ref{fig:am-e}). Values of $\epsilon$ close to unity indicate nearly circular prograde orbits in the disk plane, values near zero a dispersion-dominated distribution, and values near $-1$ a nearly perfect counter-rotating disk. This method does not rely on strict spatial cuts, reducing contamination between components, and applies to galaxies spanning a wide range of masses and sizes, though it requires some ad hoc parameters, as described below.

We use the \textsc{CIELO} database presented by \citet{gonzalez-jara+2025}, inspecting $54$ galaxies with stellar masses in the range $10^{8.0}$--$10^{10.7}\,\mathrm{M}_\odot$. Disk components consist of stellar particles with circularity $|\epsilon|\geq 0.5$, energy $E \geq E_\mathrm{bin}$, and radius $r \leq 1.5\,R_\mathrm{opt}$\footnote{The optical radius $R_{\mathrm{opt}}$ encloses 83\% of the baryonic mass of the simulated galaxy.}, where $E_\mathrm{bin}$ is the minimum energy at $r\sim 0.5\, R_\mathrm{opt}$. We focus on bulge stellar populations at $z = 0$, defined as those with binding energy smaller than $E_\mathrm{bin}$, corresponding to orbital apocentres smaller than $\sim0.5\,R_{\mathrm{opt}}$. Optical radii for the suite range within $\left[1.10, 12.30\right]\,$kpc. Since bulge classification imposes no cut in $\epsilon$, bulges can include both dispersion- and rotation-dominated (inner disk) components (see \citealt{munoz-escobar2026}; Muñoz-Escobar et al. in prep.).

The choice of \textsc{CIELO} for this study builds directly on the bulge analysis of \citet{munoz-escobar2026}, which used the same simulation suite to characterize the central structure and chemical properties of the bulge population in this sample. Continuing with \textsc{CIELO} lets us connect the properties of the stellar knots identified here to the bulge formation pathways already characterized in that work, within a single, internally consistent numerical framework, stellar population synthesis, and chemical evolution model. Beyond this continuity, \textsc{CIELO}'s spatial resolution  is sufficient to characterize the  sub-kpc central structures such as stellar knots, and its SPH-based hydrodynamical scheme produces smooth, physically well-behaved central density and metallicity profiles, without the artificial fragmentation that can otherwise complicate the interpretation of central regions. Its detailed chemical evolution model, tracking multiple elements and enrichment channels, offers a natural path toward linking the chemical signatures of knots to their formation history. Finally, the diversity of merger histories among the \textsc{CIELO} galaxies allows us to examine knots across a range of formation pathways rather than a single assembly scenario. Taken together, these properties make \textsc{CIELO} well suited to build on the bulge results of \citet{munoz-escobar2026} and to identify plausible trends worth exploring further with higher numerical resolution experiments (Acosta-Tripailao et al. in prep.)

\section{Knots in CIELO}
\label{sec:knotsincielo}
\begin{figure}[ht!]
\centering
    \includegraphics[width=\linewidth]{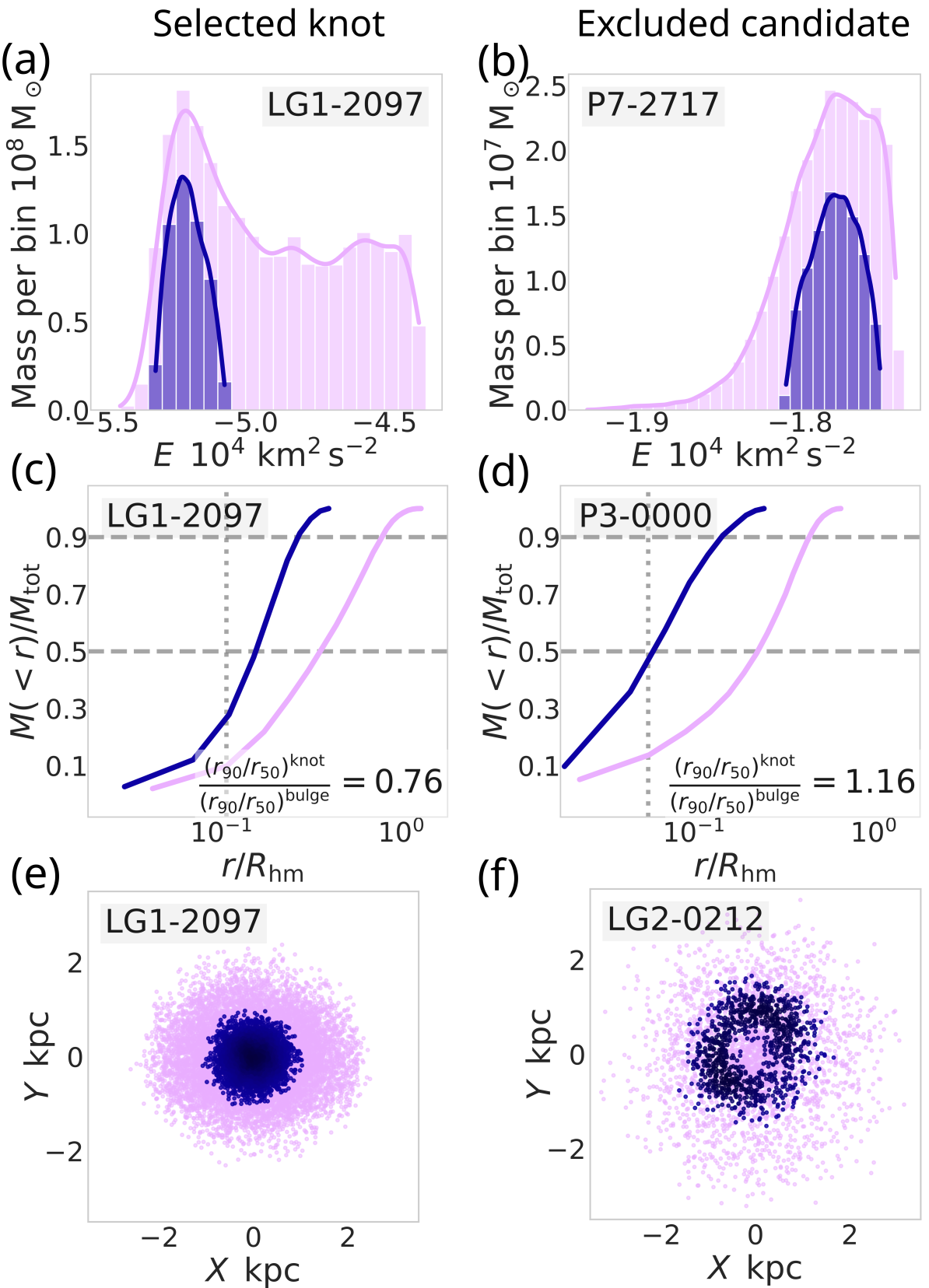}
    \caption{Examples of the three main selection criteria: kinematic \textbf{(a--b)}, concentration \textbf{(c--d)}, and morphology \textbf{(e--f)}. Left panels show a case fulfilling all criteria; right panels show discarded cases. Blue marks the knot candidate, purple the host bulge. The kinematic criterion requires the candidate's lowest-energy peak to not be shifted toward higher bulge energies. The concentration criterion requires $(r_{90}/r_{50})^{\rm knot}/(r_{90}/r_{50})^{\rm bulge} < 0.9$, seen as a steeper cumulative mass profile between horizontal dashed lines (vertical dotted lines mark $\epsilon_\mathrm{grav}$ normalized by $R_\mathrm{hm}$). The morphological criterion ensures well-defined knot structures.}
    \label{fig:criteria}
\end{figure}

The knots studied here are stellar overdensities identified at the lowest energies in the ($\epsilon, E$) space, thus tracing a concentration of mass in the deeper regions of the gravitational potential well. In this study, we require them to satisfy a set of physical criteria (see Fig.~\ref{fig:criteria}), as described below, yielding our control sample.

Candidate knots are isolated using kinematic criteria. We first apply a one-dimensional Gaussian kernel density estimation (KDE) to the energy distribution and identify the lowest-energy peak, defined as the first local maximum from the low-energy end of the distribution, as determined from the derivative of the KDE. The particles within this peak are then used to compute a two-dimensional KDE in the $(\epsilon, E)$ plane. From this distribution, we select the top $40\%$ densest regions in phase space, defining preliminary kinematic overdensities. This threshold corresponds to a percentile cut applied on the probability density distribution of stellar particles in the $(\epsilon, E)$ space, estimated via KDE. We tested a range of threshold values and adopted $40\%$ as a balance between capturing the core of the overdensity and retaining a sufficient number of particles. For both the one- and two-dimensional distributions, we use a Gaussian KDE implemented in SciPy~\citep{virtanen2020scipy}. The bandwidth choices reflect the distinct roles of each KDE: the one-dimensional $(E)$ KDE adopts Scott's rule, providing automatic sample-size-dependent smoothing suited for peak identification, while the two-dimensional $(\epsilon, E)$ KDE uses a fixed bandwidth scaling of $0.5$ to ensure consistent smoothing across systems with widely varying particle numbers. In both cases, particle masses are used as weights.

Our purposes require the selected kinematic overdensity to be anchored at the most bound end of the energy distribution; five candidate knots in which there is a clear peak at low energy, but it is not quite a the lowest boundary, were discarded (see Fig.~\ref{fig:criteria}a,b for a comparative example).

To ensure statistical robustness, knots are required to contain a minimum of $500$ stellar particles\footnote{Our final control sample spans particle numbers from $522$ up to $\sim1.7\times10^{5}$, with over $75\%$ containing more than $1000$ particles.}. For systems with fewer particles, the density threshold was adaptively increased up to $90\%$; eight candidates that remain poorly populated were discarded.

We then applied a concentration criterion, requiring the knot candidates to have a more concentrated radial mass growth than that of their host bulge. This is quantified using the ratio $\left(r_{90}/r_{50}\right)$\footnote{In this work, $r_{90}$ and $r_{50}$ are the radii enclosing $90\%$ and $50\%$ of the stellar mass of a given stellar population, respectively.}, commonly employed in studies of galaxy morphology to quantify how centrally concentrated the stellar light or stellar mass distribution is. We define a normalized concentration parameter as $\frac{\left(r_{90}/r_{50}\right)^{\rm knot}}{\left(r_{90}/r_{50}\right)^{\rm bulge}}$, where values approaching zero indicate increasingly compact knots relative to the bulge, values close to unity correspond to comparable concentrations, and values larger than unity indicate knots that are less concentrated than their host bulge. Only candidates satisfying a ratio smaller than $0.9$ are retained, which ensures that the knot is more centrally concentrated than the bulge as a whole. Seven knot candidates were discarded by this criterion (see Fig.~\ref{fig:criteria}c,d for a comparative example, where cumulative mass profiles are shown as a function of radius normalized by $R_\mathrm{hm}$\footnote{$R_\mathrm{hm}$ is the stellar half-mass radius of the simulated galaxy.}).

A qualitative morphological inspection is performed to retain only centrally concentrated knots structures and exclude ring-like or hollow structures. Three candidates are discarded under this criterion (see Fig.~\ref{fig:criteria}e,f for a comparative example).

Finally, we retain only those knot candidates whose sizes $r_{90}>\epsilon_\mathrm{grav}$ in their respective resolution runs. This criterion leads to the rejection of two candidates. Additionally, a resolution assessment of the identified knot population is presented in Appendix~\ref{sec:appendix_reseff}.

Through these different levels of selection, the initial sample of 54 knot candidates was reduced to a final set of 28 robust \textsc{CIELO} knots ($52\%$), which defines our control sample and is the focus of the remainder of this work. We note that among the 28 selected knot-hosting galaxies only P7-2627 hosts a bar, while 4 of the 26 discarded candidates also host a bar. The interplay between knots and bars will be addressed in a companion paper (Acosta-Tripailao et al., in prep.).

We note that the final knot sample may depend not only on numerical resolution (see Appendix~\ref{sec:appendix_reseff}) but also on the adopted sequence of selection criteria. We verified, however, that the trends reported below are already present and consistent in the full sample of 54 knot candidates prior to applying all the selection criteria. Considering the full sample results in substantially larger scatter and more extended percentile ranges, reflecting the presence of lower-fidelity candidates; our selection criteria reduce this scatter, isolating a cleaner signal without altering the qualitative behavior of the underlying trends.

\section{Characterization of knots}
\label{sec:charact}
Once the knots were dynamically isolated, we characterize their chemical enrichment, formation timescales, birth locations, and spatial distributions relative to the surrounding bulge stellar populations. For this comparison, we defined two reference components, excluding knot stars from both: (i) a co-spatial component, consisting of bulge stars in the same spatial region as the knot but not belonging to it, e.g., stars on eccentric orbits that traverse the inner region with higher energy; and (ii) the rest of the bulge, which includes the co-spatial component. Note that component (ii) is a superset of component (i), allowing for both a local and a global comparison with the surrounding bulge. The comparisons reported throughout this section should be considered in light of the fact that 8 of our 28 knots ($29\%$) have $r_{90} < 2\epsilon_\mathrm{grav}$. At these scales, numerical resolution may affect the central density distribution and dynamical separation between components. Stars near the knot-bulge boundary may therefore be classified differently depending on it (see Appendix~\ref{sec:appendix_reseff}).

\subsection{Alpha-element distribution}
\begin{figure*}
    \centering
    \includegraphics[width=.8\linewidth]{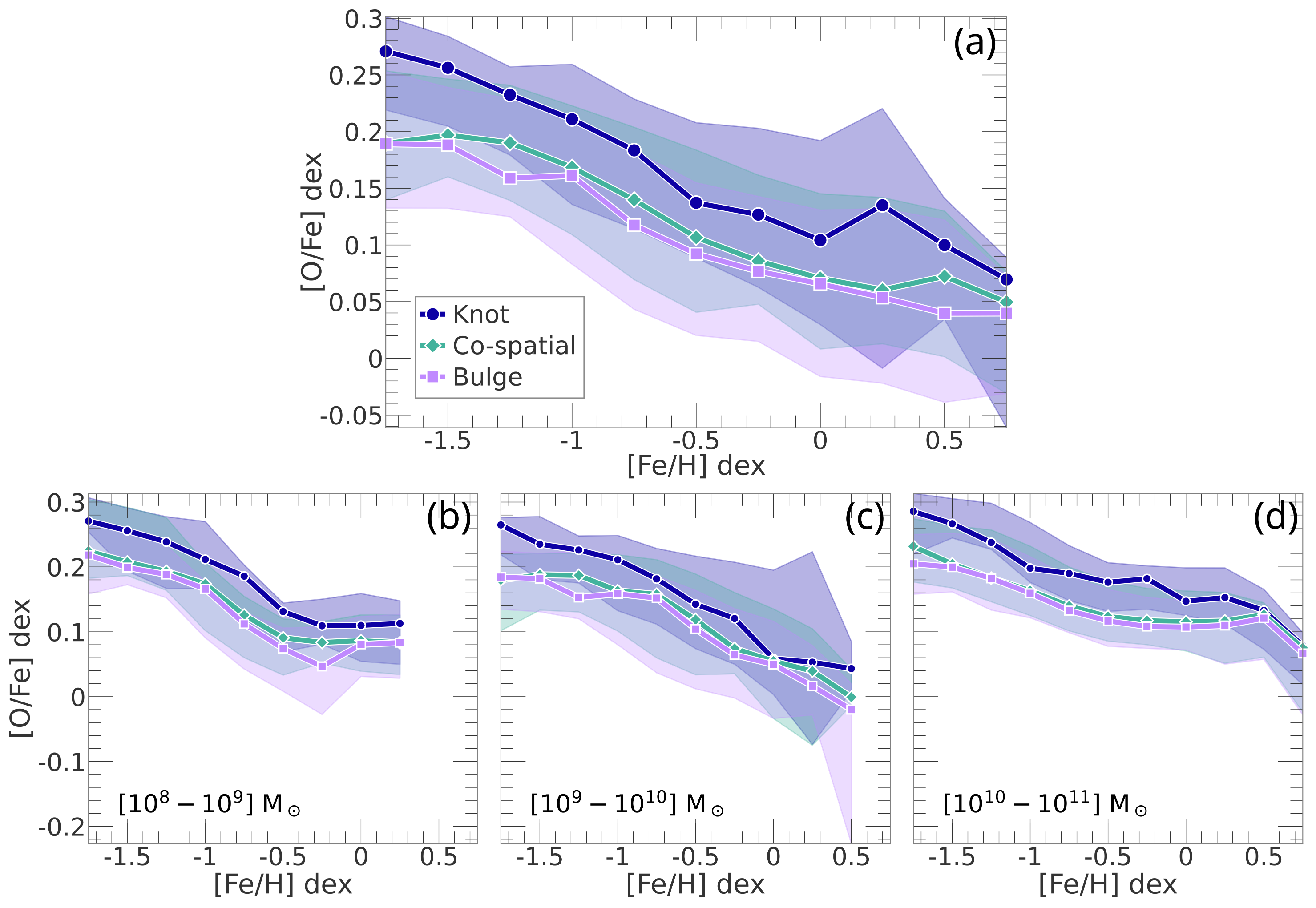}
    \caption{Panel \textbf{(a)}: Stacked $\left[\mathrm{O}/\mathrm{Fe}\right]$–$\left[\mathrm{Fe}/\mathrm{H}\right]$ relations based on the mass-weighted median across galaxies. Solid curves for the knot (blue), co-spatial (green), and bulge (purple) stellar populations. Shaded regions show the 16th–84th percentiles. Panels \textbf{(b--d)}: Same as panel (a), but grouped by stellar mass of the corresponding host galaxies, as indicated by the inset labels.}
    \label{fig:alphaplane_median}
\end{figure*}

We estimated the individual alpha-element abundance planes of the stellar populations corresponding to the 28 knots (see Fig.~\ref{fig:individual_alphaplanes}), the co-spatial stellar populations and the corresponding bulges. This plane stores relevant information on the star formation history of the stellar populations. The alpha-elements are injected promptly by core-collapse SNe, while iron comes from both sources, core-collapse SNe and Type Ia SNe, building up more gradually due to the delayed contribution from Type Ia SNe. Therefore, this plane can be used as a chemical clock, roughly evolving from high $\left[\alpha/\mathrm{Fe}\right]$, low $\left[\mathrm{Fe}/\mathrm{H}\right]$ to low $\left[\alpha/\mathrm{Fe}\right]$, high $\left[\mathrm{Fe}/\mathrm{H}\right]$ \citep[e.g.][]{Tinsley1979,McWilliam1997,iza+2025}.

Figure~\ref{fig:alphaplane_median}a presents the stacked $\left[\mathrm{O}/\mathrm{Fe}\right]$–$\left[\mathrm{Fe}/\mathrm{H}\right]$ relations for each defined stellar component (knots, co-spatial components, and bulges). These are constructed as mass-weighted median abundance profiles in fixed $\left[\mathrm{Fe}/\mathrm{H}\right]$ bins, following the bootstrap procedure described in Appendix~\ref{sec:appendix_alphaplane}. The stacked relation in each bin is then defined as the median $\left[\mathrm{O}/\mathrm{Fe}\right]$ across all galaxies, and the shaded regions represent the 16th and 84th percentiles of the distribution. A systematic offset in $\left[\mathrm{O}/\mathrm{Fe}\right]$ is observed among the different components: the knot population (blue lines) is shifted toward higher alpha-element enhancement relative to both the co-spatial and bulge components. The mean offset is approximately $0.05\,$dex. 

Figure~\ref{fig:alphaplane_median}b--d shows the same chemical plane, with the stacking performed by grouping systems according to the  stellar mass of the corresponding host galaxy. Across all galaxy mass, the knots remain systematically more alpha-enhanced than both the co-spatial and bulge components, indicating that this behavior is independent of galaxy mass, though the offset is larger in the high-mass bin, and less pronounced in low-mass galaxies. An analogous stacking is performed for the mode in Fig.~\ref{fig:alphaplane_mode} (see Appendix~\ref{sec:appendix_alphaplane}).

Notably, the systematic alpha-element enhancement persists across all analyzed mass ranges. The enhanced $\left[\alpha/\mathrm{Fe}\right]$ ratios of the knots suggest that they may have followed different evolutionary pathways compared to the surrounding co-spatial and bulge populations. Higher $\left[\alpha/\mathrm{Fe}\right]$ ratio is consistent with shorter star formation timescales, during which Type~Ia supernovae had insufficient time to significantly enrich the ISM with iron. This would suggest that knots experienced a more rapid formation compared to the more extended star formation history of the surrounding co-spatial stellar component. In the following subsection, we explore this interpretation further by analyzing the formation timescales of the different stellar components. 

\subsection{When and how fast did knots form?}
\begin{figure}[ht!]
   \centering
   \includegraphics[width=\hsize]{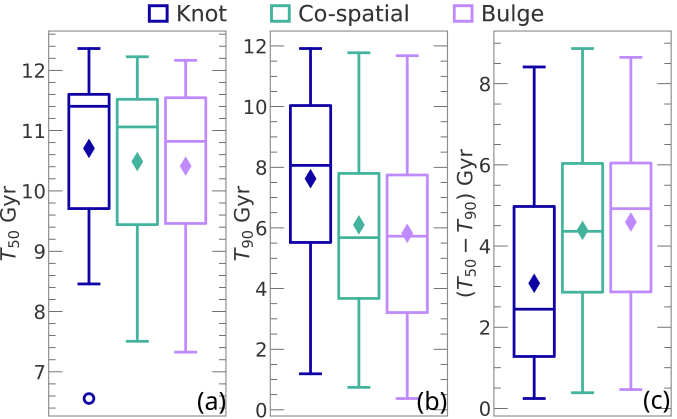}
     \caption{Panels \textbf{(a)} and \textbf{(b)} show the distributions of lookback times at which a component formed $50\%$ and $90\%$ of its stellar mass, respectively; panel \textbf{(c)} shows the difference between these timescales, indicating how rapidly each structure formed. In each panel, knots are shown at the left (blue), co-spatial structures in the middle (green), and bulges at the right (purple). Boxplots show the mean (diamond), median (horizontal line), and interquartile range ($25$–$75\%$). Whiskers extend to the most extreme points within $1.5\times$IQR; values beyond this range are shown as outliers (circles). See Table~\ref{table:t5090_stats} for exact values.}
    \label{fig:t5090}
\end{figure}

A useful way to characterize the timescales of stellar mass assembly during the evolution of a galaxy or a given structure is through cumulative formation time statistics such as $T_{50}$ and $T_{90}$. These quantities correspond to the lookback times at which $50\%$ and $90\%$ of the final stellar mass in a given component have been formed, respectively, as inferred from the ages of the present-day stellar populations.

Figure~\ref{fig:t5090}a,b shows the distributions of $T_{50}$ and $T_{90}$ expressed as boxplots (the boxplot conventions are described in the figure caption and the exact values are displayed in Table~\ref{table:t5090_stats}). While the $T_{50}$ distributions are quite similar among the different components, with their medians marginally decreasing across samples, there is a systematic decrease of $T_{90}$ from the knot populations to the bulge's ones accompanied by an increase in the dispersion. The $T_{90}$ distributions reveal that knots reach $90\%$ of their final stellar mass $\sim$2$\,$Gyr earlier than both the co-spatial component and the bulge. Specifically, knots exhibit a median $T_{90}$ of $\sim$8$\,$Gyr ago, compared to $\sim$6$\,$Gyr ago for the co-spatial component and the bulge. We acknowledge that the variety of the $T_{90}$ is appreciably larger than $T_{50}$, suggesting a diversity of star formation history for the intermediate and young stellar populations. Nevertheless, the statistical trend for the earlier history of formation of stars in the knots is present.

These results indicate that knots, while starting to form stars at the same epochs as the reference components, completed their star formation at earlier epochs, pointing to a more rapid star formation history compared to the more extended formation of the co-spatial component. Figure~\ref{fig:t5090}c further illustrates this behavior through the distribution of $T_{90}-T_{50}$, which traces the characteristic formation timescale (exact values are displayed in Table~\ref{table:t5090_stats}). Knots show systematically smaller values of $T_{90}-T_{50}$, confirming that they form on shorter timescales, with a median of $\sim$2$\,$Gyr, while the remaining stellar population in the bulge shows a median timescales of $\sim$5$\,$Gyr. This finding is consistent with knot populations exhibiting alpha-element enhancement as discussed in Fig.~\ref{fig:alphaplane_median}.

\subsection{How did knots form?}
   \begin{figure*}
        \centering
        \includegraphics[width=\linewidth]{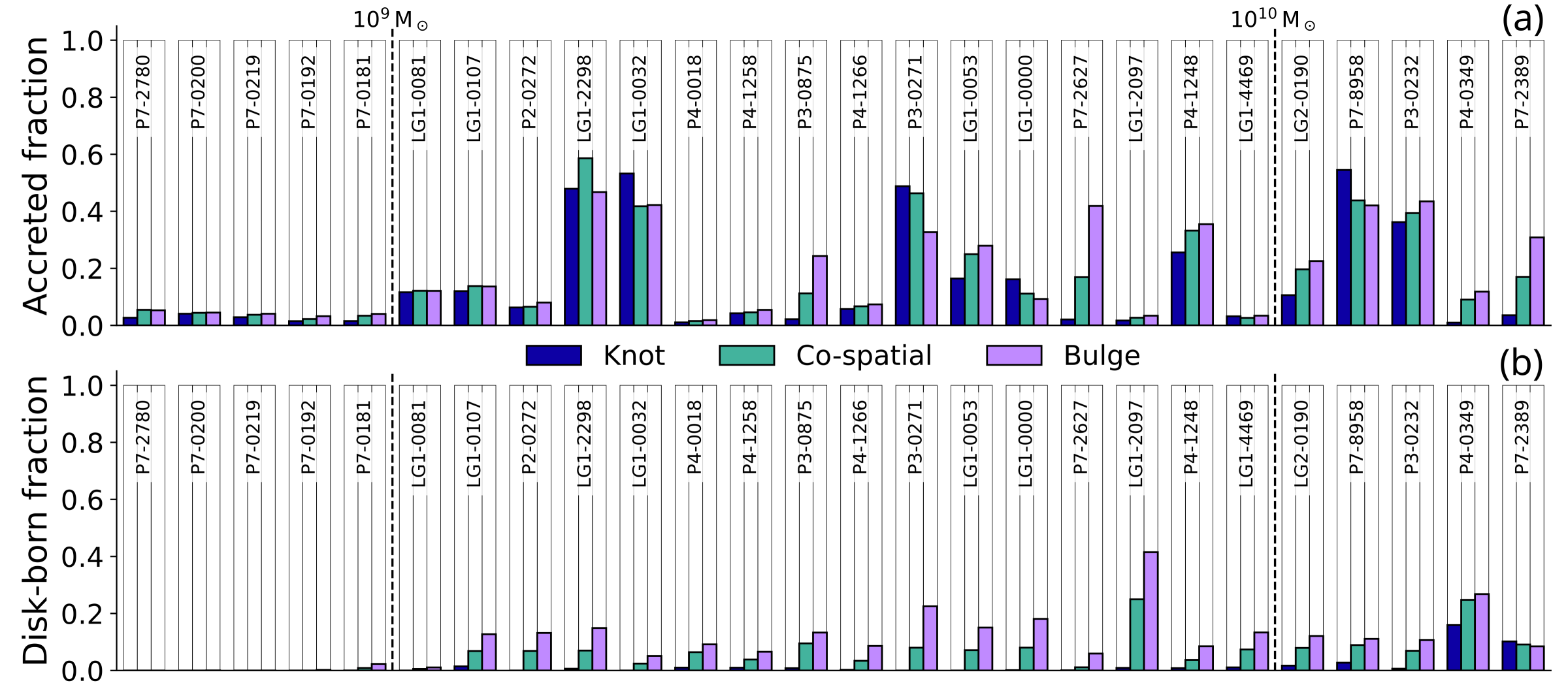}
    \caption{Bar plots show the \textbf{(a)} accreted and \textbf{(b)} disk-born mass fractions for each galaxy, with galaxy labels at the top. In each galaxy, knots are shown at the left (blue), co-spatial structures in the middle (green), and bulges at the right (purple). Galaxies are ordered by increasing stellar mass, with vertical dashed lines separating three mass regimes.}
        \label{fig:accreted_disk_individual}
    \end{figure*}

\begin{figure}[ht!]
   \centering
   \includegraphics[width=1\hsize]{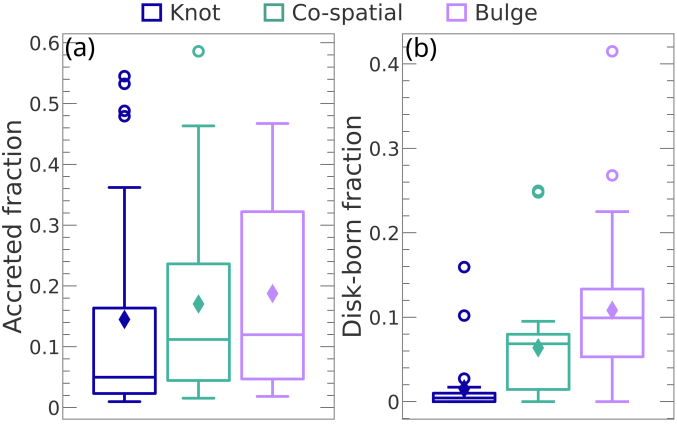}
    \caption{\textbf{(a)} Boxplot of the accreted mass fraction combining all galaxies shown in Fig.~\ref{fig:accreted_disk_individual}a. \textbf{(b)} Boxplot of the disk-born mass fraction combining all galaxies shown in Fig.~\ref{fig:accreted_disk_individual}b. In each panel, knots are shown at the left (blue), co-spatial structures in the middle (green), and bulges at the right (purple). Boxplots show the mean (diamond), median (horizontal line), and interquartile range ($25$–$75\%$). Whiskers extend to the most extreme points within $1.5\times$IQR; values beyond this range are shown as outliers (circles). See Table~\ref{table:acc_disk_fraction_stats} for exact values.}
    \label{fig:acc_disk_boxplots}
\end{figure}

The stellar populations that make up galactic bulges can arise through different formation channels, a diversity that is also seen in the \textsc{CIELO} simulations~\citep[][see also Casanueva et al., in prep.]{munoz-escobar2026}. Then, the previous results naturally raise the question of the origin of the stellar populations in the knots: whether they formed in-situ in the progenitor bulge or in the disk component and then migrated, or were accreted through interactions and mergers~\citep{romano+2023}. In latter cases, they ended up as structures of accreted systems embedded in the heart of galaxies. To address this question, we tracked all the bulge stellar populations to back to their site of formation. We analyzed the stellar populations associated with the knots by separating them into in-situ and accreted components~\citep{gonzalez-jara+2025}. The in-situ population comprises stars bound to the central galaxy at the first available snapshot after their formation. The accreted population includes stars formed in satellite galaxies, regardless of whether they were inside or outside the virial radius. The accreted mass fractions for the knots, co-spatial and bulge components in each galaxy are shown in Fig.~\ref{fig:accreted_disk_individual}a.

In particular cases, the merger tree algorithm may misidentify the main progenitor during high-redshift mergers of similar mass, leading to in-situ stars being incorrectly classified as accreted. To avoid this, systems with more than $50\%$ of bulge stars classified as accreted are excluded; this occurs only twice in our sample ($M_\mathrm{gal} = 2.51 \times 10^9$ and $4.09 \times 10^{10}\;\mathrm{M}_\odot$).

Within the in-situ population, we further classify stellar populations according to their site of formation within the galaxies. This allows us to assess whether the stellar population migrated inward from the disk or were formed directly in the bulge. Stars are classified as disk-born if they formed within the disk component, as identified by the AM-E method (see Section~\ref{sec:CIELO}). The disk-born mass fraction of the knot, co-spatial, and bulge components in each galaxy is shown in Fig.~\ref{fig:accreted_disk_individual}b. In low-mass galaxies (e.g., P7-2780, P7-0200, P7-0219, P7-0192, with stellar masses $\lesssim 10^{9}\,\mathrm{M}_\odot$), there is little evidence for significant disk migration in general. These galaxies were simulated with high numerical resolution ($\sim10^{4} \,\mathrm{M}_\odot$), but the number of particles is lower than for more massive systems and could be more affected by numerical  effects. Hence, more systems and, probably higher resolutions, are needed to draw a robust conclusion. For more massive galaxies ($\gtrsim 10^{9}\,\mathrm{M}_\odot$), whereas disk migration contributes to the co-spatial component and the host bulge with median mass fractions of about $10\%$, the knots exhibit only a negligible contribution from disk-born stellar populations. Only the two most massive galaxies in this subsample, P4-0349 ($1.84\times10^{10}\,\mathrm{M}_\odot$) and P7-2389 ($5.43\times10^{10}\,\mathrm{M}_\odot$), have non-negligible contribution of disk-born stars to the knots, with a $16\%$ and $10\%$, respectively.

Figure~\ref{fig:acc_disk_boxplots} summarizes the statistics presented in Fig.~\ref{fig:accreted_disk_individual} using boxplots (the boxplot conventions are described in the figure caption and the exact values are displayed in Table~\ref{table:acc_disk_fraction_stats}). In Fig.~\ref{fig:acc_disk_boxplots}a we find that stellar populations forming the knots exhibit a median accreted mass fraction of $5\%$, approximately half that of both the co-spatial component ($11\%$) and the host bulge ($12\%$). However, the distributions show substantial scatter, which increases systematically from knots to the co-spatial component and the host bulge. An inspection of individual cases in Fig.~\ref{fig:accreted_disk_individual}a indicates that when one component exhibits a significant accreted contribution, the other components tend to do so as well, thus, this trend does not provide a fully conclusive distinction between the different stellar structures. In contrast, Fig.~\ref{fig:acc_disk_boxplots}b evidences a clear difference between the components. While the co-spatial components and the host bulges exhibit small but non-negligible contributions from disk-born stars (medians of $7\%$ and $10\%$, respectively), the knots show almost no such contribution (median $\sim$0$\%$), with the exception of a few cases in the most massive galaxies, P4-0349 and P7-2389 (Fig.~\ref{fig:accreted_disk_individual}b), consistent with the trend noted above. This could suggest that in more massive systems, migration may contribute to the formation of the knot. Indeed, although the disk-born fraction is not dominant, its presence points to a more complex formation history in MW-mass galaxies, where secular processes play a more significant role. We will discuss the effect of radial migration on the knot formation in MW-mass galaxies in a future paper. 

\subsection{Where did knots acquire gas from?}
\begin{figure*}
    \centering
    \includegraphics[width=\linewidth]{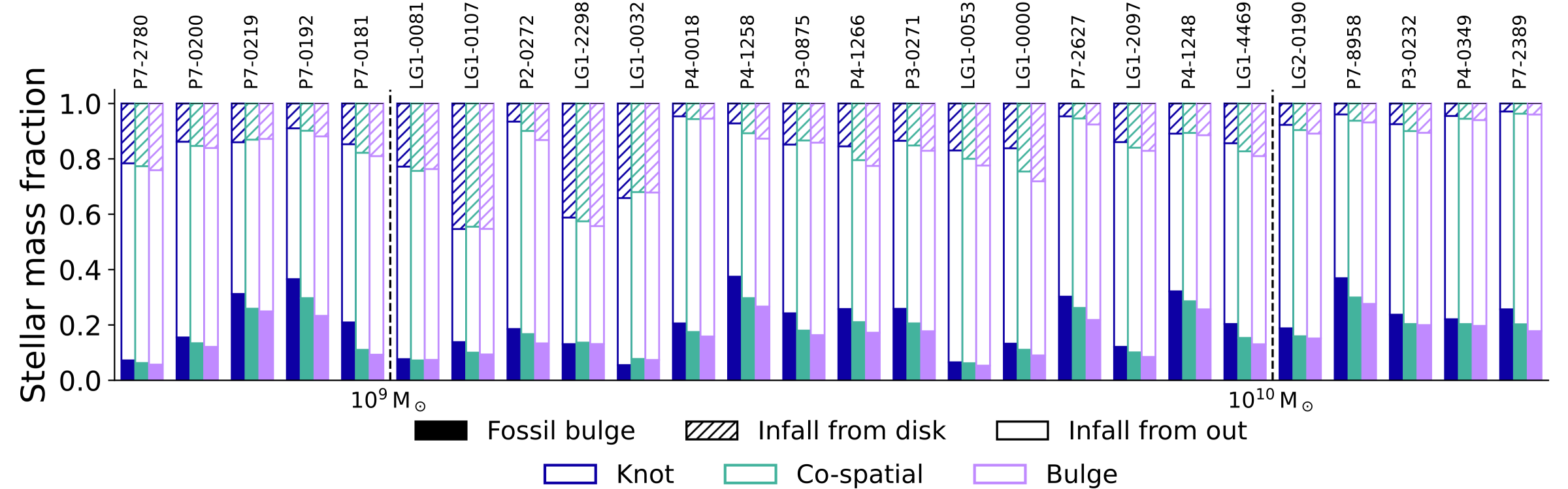}
    \caption{Bar plots showing the stellar mass fractions according to the gas origin of in-situ stellar particles for each galaxy, with galaxy labels at the top. Within each galaxy, knots are shown on the left (blue), co-spatial structures in the middle (green), and bulges on the right (purple). Bar patterns indicate gas origin: solid for fossil bulge gas, hatched for disk infall, and unfilled for infall from outside the disk and bulge. Galaxies are ordered by increasing stellar mass, with vertical dashed lines separating three mass regimes.}
    \label{fig:gas_origin_individual}
\end{figure*}

\begin{figure}
    \centering
    \includegraphics[width=\linewidth]{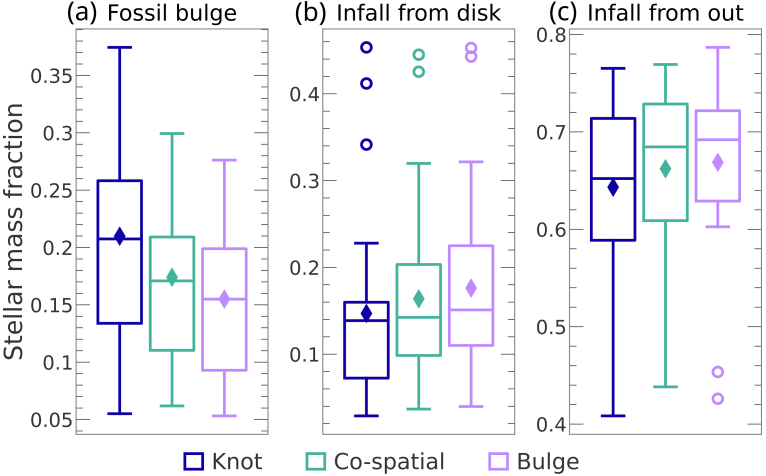}
    \caption{Boxplots of the in-situ stellar mass fraction according to gas assembly channel, combining all galaxies shown in Fig.~\ref{fig:gas_origin_individual}: \textbf{(a)} fossil bulge, \textbf{(b)} disk infall, and \textbf{(c)} external infall. In each panel, knots are shown on the left (blue), co-spatial structures in the middle (green), and bulges on the right (purple). Boxplots show the mean (diamond), median (horizontal line), and interquartile range ($25$--$75\%$). Whiskers extend to the most extreme points within $1.5\times$IQR; values beyond this range are shown as outliers (circles). See Table~\ref{table:gas_origin_stellar_fractions} for exact values.}
    \label{fig:gas_origin_boxplots}
\end{figure}

\begin{figure}
    \centering
    \includegraphics[width=\linewidth]{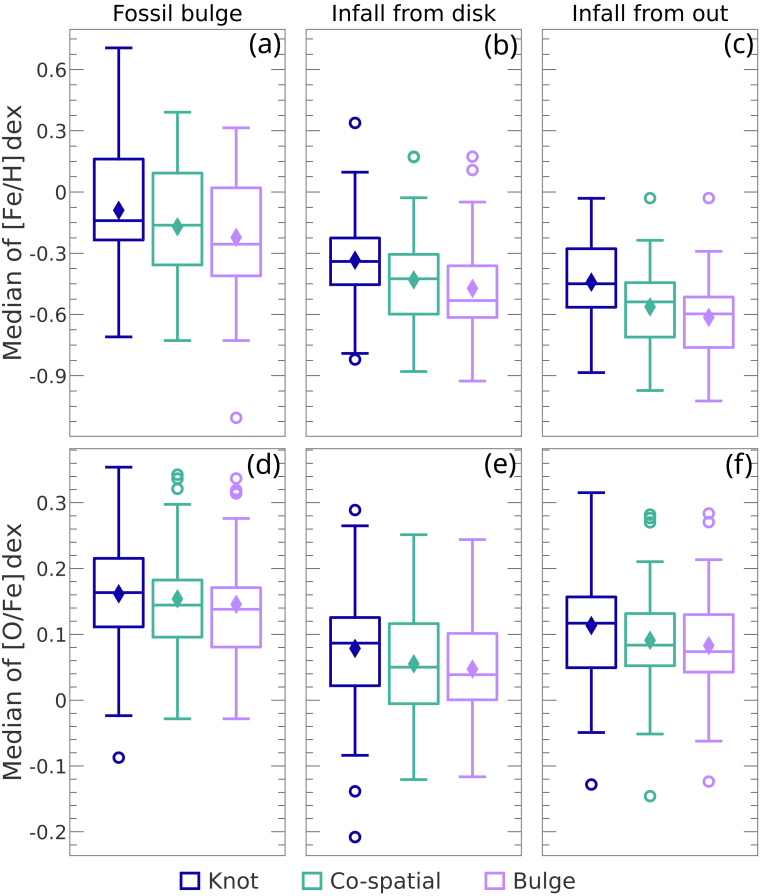}
    \caption{Boxplots of the mass-weighted median $\left[\mathrm{Fe}/\mathrm{H}\right]$ \textbf{(a--c)} and $\left[\mathrm{O}/\mathrm{Fe}\right]$ \textbf{(d--f)} for stellar populations from each assembly channel shown in Fig.~\ref{fig:gas_origin_boxplots}. See Tables~\ref{table:star_feh_stats} and \ref{table:star_ofe_stats} for exact values.}
    \label{fig:stellar_gaschannel_FeHOFe_boxplots}
\end{figure}

To characterize the gas that forms the in-situ stellar populations, we adopt a tracking approach in which we identify the progenitor gas particles associated with each in-situ stellar particle and follow them across time. At each  available snapshot, the spatial position of each gas particle is classified relative to the main galaxy by defining  three regions as follows: a bulge region as a sphere of radius $0.5\,R_{\mathrm{opt}}$, a disk region as a rectangular box of side $1.5\,R_{\mathrm{opt}}$ and height $0.5\,R_{\mathrm{hm}}$, excluding the bulge region, and an outer region corresponding to the remaining volume of the simulation. For each gas particle we accumulate the time spent in each region weighted by its gas mass, yielding the total fractions of mass-time spent in the bulge $f_{\rm bulge}$, disk $f_{\rm disk}$, and outer regions $f_{\rm out}$; i.e., a fraction of $0.5$ means that half of the gas particle's mass-weighted time was spent in the respective region. Based on these fractions, the gas is classified into three channels following a hierarchical classification scheme, with two main objectives: first, to identify gas that remained confined to the bulge region throughout its history, hereafter, fossil bulge, $f_{\rm bulge}\ge0.9$. Secondly, to determine the origin of the remaining gas, distinguishing between disk infall ($f_{\rm disk}\ge0.05$) and external infall ($f_{\rm out}\ge0.2$ and $f_{\rm disk}<0.05$). In Fig.~\ref{fig:gas_origin_individual} we show the mass fraction of each stellar population formed from the different gas pathways; cases not meeting any of the above criteria represent a negligible contribution to the stellar mass fractions across all components and galaxies ($< 0.5\%$), and are therefore folded into the external infall channel. The combined distributions across the full galaxy sample are presented as boxplots in Fig.~\ref{fig:gas_origin_boxplots} (the boxplot conventions are described in the figure caption and the exact values are displayed in Table~\ref{table:gas_origin_stellar_fractions}).

Figure~\ref{fig:gas_origin_boxplots}a displays the fossil bulge mass fractions across all components, with knots showing the largest contribution, reaching a median of $21\%$ (and maximum values up to $37\%$), compared to $17\%$ and $16\%$ for co-spatial and bulge components, respectively. As shown in Fig.~\ref{fig:gas_origin_boxplots}b, all three components show comparable disk infall mass fractions, with medians of $14\%$, $14\%$, and $15\%$ for knots, co-spatial, and bulge, respectively. For the knots, these non-negligible contributions suggests that a primordial disk component contributed gas prior to the knot formation epoch. Finally, Fig.~\ref{fig:gas_origin_boxplots}c reveals considerable external infall mass fractions across all components, indicating the presence of a gas subcomponent that fell directly into the central regions, possibly from the CGM or through cold filamentary inflows. Knots show a slightly lower external infall contribution than the other components, with a median of $65\%$ compared to $68\%$ and $69\%$ for co-spatial and bulge, respectively. In summary, the gas tracking analysis demonstrates that in-situ stellar populations of knots are assembled from three main channels, with external infall being the largest, fossil bulge gas the second, and disk infall the smallest. The largest difference among components is found in the fossil bulge fraction, where knots show the highest median contribution, consistent with their nature as early-formed, centrally confined structures.

Figures~\ref{fig:acc_disk_boxplots} and \ref{fig:gas_origin_boxplots} altogether indicate that simulated knots show a larger contribution of stellar populations formed from gas that fell quickly into the central regions and gas that remained confined to the central component for a longer time, compared to both co-spatial and bulge components, and may therefore be interpreted as fossil bulge structures of the early stages of galaxy evolution. By contrast, both the bulge component (defined here as the host bulge excluding the knot) and the co-spatial component display stronger signatures of stellar disk migration and lower contribution of fossil bulge gas. Consequently, if the co-spatial and knot populations exhibit distinct chemical evolution patterns (see Fig.~\ref{fig:alphaplane_median}), this difference is in part driven by the stronger contribution of disk-migrated stars and gas to the co-spatial component, rather than by stellar migration into the knot. This adds to their clear different star formation histories discussed above.

We further characterized the metallicity of these stellar populations by computing the mass-weighted median of $\left[\mathrm{Fe}/\mathrm{H}\right]$ and $\left[\mathrm{O}/\mathrm{Fe}\right]$ per gas assembly channel. Figure~\ref{fig:stellar_gaschannel_FeHOFe_boxplots} shows the distributions of per-component medians across the galaxy sample, whose exact values are reported in Tables~\ref{table:star_feh_stats} and \ref{table:star_ofe_stats}. Although differences among components are modest, with knots being the most enriched in both iron and alpha-element abundances, strong differences emerge among formation channels. From Fig.~\ref{fig:stellar_gaschannel_FeHOFe_boxplots}a--c, stars with a fossil bulge origin are the most metal-enriched, having spent more time in the bulge where early, intense star formation, likely a central starburst, drove significant enrichment. In contrast, stars born from external infall are the least enriched, with disk infall gas lying between the two. Fig.~\ref{fig:stellar_gaschannel_FeHOFe_boxplots}d--f suggests that the fossil bulge channel plays a leading role in the alpha-element enhancement of the knots. Unlike the iron abundance case, however, the disk infall channel no longer occupies an intermediate position in terms of enrichment, showing slightly lower alpha-element abundances than the external infall channel. This is expected, as the disk sustains a more continuous star formation rate, lacking the intense early burst characteristic of the bulge. These results are particularly relevant in light of the alpha-element enhancement found in knot stars (see Fig.~\ref{fig:alphaplane_median}): they support the interpretation that fossil bulge gas is the primary channel associated with the alpha-element enhancement of knots, consistent with an early, in-situ enrichment history. In this context, a deeper gravitational potential well could diminish gas ejection and enhance gas retention, facilitating the build-up of a massive and centrally concentrated stellar component such as a knot, sustaining the prolonged enrichment needed to produce this distinct chemical signature.

\subsection{How are morphologies of the knots?}
\begin{figure}[ht!]
    \centering
    \includegraphics[width=\linewidth]{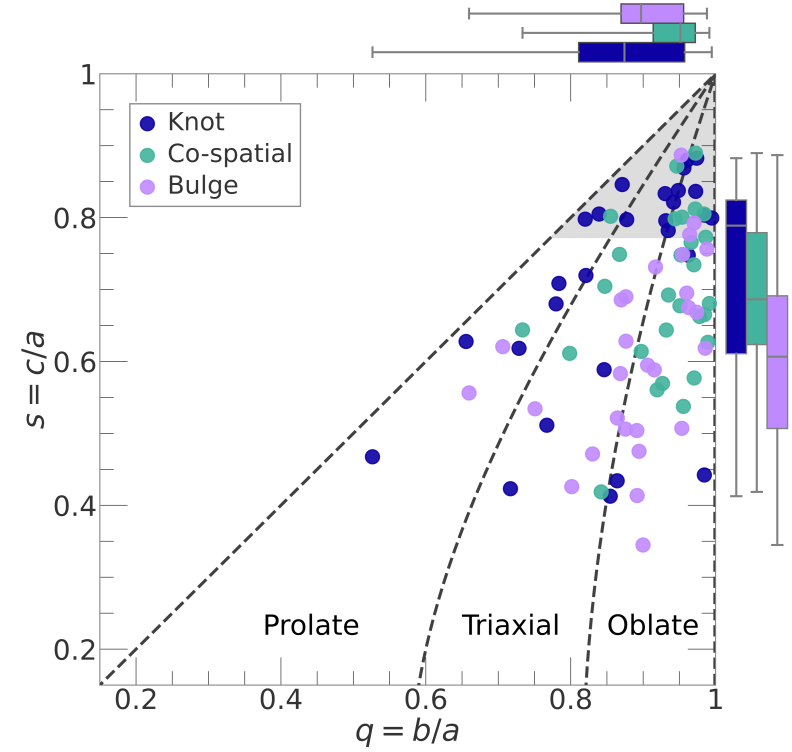}
    \caption{Contour plot of the triaxiality parameter $T$ as a function of $q$ and $s$, distinguishing oblate/disk-like ($0 \leq T < 1/3$), triaxial ($1/3 \leq T < 2/3$), and prolate ($2/3 \leq T \leq 1$) shapes (dashed lines); the diagonal ($q=s$) and vertical ($q=1$) dashed lines mark perfectly prolate and oblate objects, respectively. The gray region shows the nearly spherical regime ($s, q \approx 1$), where $T$ becomes ill-defined. Points show knots (blue), co-spatial structures (green), and bulges (purple); boxplots give the median, interquartile range, and min--max whiskers for each component. Contours plot is based on Fig.~11 from \citet{Keith_2025}. See Table~\ref{table:triaxi_regimes} for population counts per regime.}
    \label{fig:triax}
\end{figure}
To morphologically characterize the knots, we use the mass quadrupole tensor $S_{jk} = \frac{1}{M} \sum_{i} m_i \, r_{i,j} \, r_{i,k}$, where $m_i$ and $r_{i,j}$ are the mass and position (relative to the galaxy center) of the $i$-th particle, and $M$ is the total mass of the component. Diagonalizing this tensor yields eigenvalues $\lambda_a \geq \lambda_b \geq \lambda_c$, defining axes $a=\sqrt{\lambda_a}$, $b=\sqrt{\lambda_b}$, $c=\sqrt{\lambda_c}$, and axis ratios $s=c/a$, $q=b/a$. These define the triaxiality parameter $T = (1 - q^2)/(1 - s^2)$, where $0 \leq T < 1/3$ indicates oblate/disk-like shapes, $1/3 \leq T < 2/3$ triaxial, and $2/3 \leq T \leq 1$ prolate. For nearly spherical objects ($s$, $q \approx 1$), $T$ becomes ill-defined due to numerical instability from small changes in axis ratios. This approach has been applied to study the morphology of astrophysical structures, such as dark matter halos and/or galaxies~\citep[e.g.,][]{tissera_1998,Thob+19,Casanueva+2022,cataldi_2023,valenzuela_remus2024,Keith_2025}

Figure~\ref{fig:triax}, showing the two triaxiality parameters for knots and reference samples, reveals several interesting features. First, knots exhibit a wide range of morphologies, including oblate ($14\%$), triaxial ($11\%$), and prolate ($21\%$) shapes, unlike the co-spatial and bulge components, which are mainly oblate. Most importantly, the bulk of knots ($54\%$) is located in the nearly spherical regime ($s, q \approx 1$), where the triaxiality parameter becomes ill-defined. This feature indicates that we are likely identifying in-situ spheroids among the knots. This regime broadly overlaps with the spherical classification adopted in other studies of simulated bulges~\citep[e.g.,][]{gargiulo+2019, costantin+2018b}, who also treat near-spherical shapes as a separate regime, although we interpret this result with due caution given the numerical instability of $T$ in this limit. Second, the fact that the distributions of $s$ and $q$ differ between components indicates that some knots do not inherit the morphology of their host bulge, nor do they share it with their surroundings (this is verified in Appendix~\ref{sec:appendix_triaxiality}). Thus, knots are structures mostly independent of both their host bulge and co-spatial stellar environment.

\section{Discussion}
\label{sec:discu}
The potential existence of a mass concentration in the center of galaxies has its precedents in our own Galaxy. A central knot in the MW was first reported by \citet{horta+2025_knot} and \citet{rix+2024_mrichknot}. However, the \textsc{CIELO} sample spans a much wider range of galaxy masses than the MW, with only its most massive galaxies reaching the MW stellar mass range; consequently, many \textsc{CIELO} knots, hosted by lower-mass galaxies, are smaller than the reported MW knot. Together with our distinct selection approach, these structures should therefore not be regarded as direct counterparts of the reported MW knot, but rather as evidence for a more general class of centrally concentrated stellar overdensities hosted by galaxies with a variety of stellar masses. In the following, we discuss their findings in the context of our results, along with caveats regarding our enrichment prescriptions.

\subsection{Comparison with observations}
\begin{figure}[ht!]
   \centering
   \includegraphics[width=\linewidth]{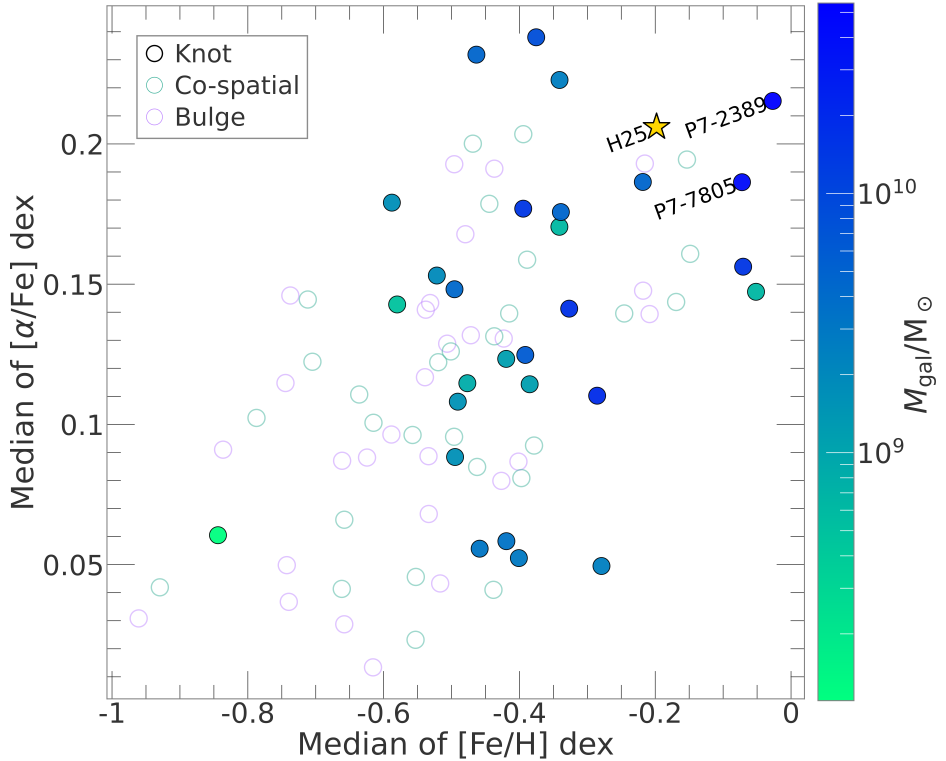}
    \caption{Mass-weighted median $\left[\alpha/\mathrm{Fe}\right]$ vs. $\left[\mathrm{Fe}/\mathrm{H}\right]$ at $z = 0$ for \textsc{CIELO} galaxies, compared with the MW knot estimated from \citet{horta+2025_knot} (H25; yellow star). Two knots belonging to galaxies in the MW-mass range are indicated. Each point represents a component: knots are color-coded by $M_\mathrm{gal}$, while co-spatial structures and bulges are shown as open circles with green and purple edges, respectively.}
    \label{fig:horta}
\end{figure}

\citet{horta+2025_knot} performed a dynamical disentangling of the disk, bar, and knot among red giant stars from the APOGEE--Gaia surveys within a galactocentric radius of $r<5\,\mathrm{kpc}$ and with $\left[\mathrm{Fe}/\mathrm{H}\right] > -0.8$. Their method consisted of an angular-momentum decomposition using a Gaussian mixture model (GMM) likelihood. From the dataset, kindly provided by Danny Horta, we estimated the median $\left[\mathrm{Fe}/\mathrm{H}\right]$ and $\left[\mathrm{Mg}/\mathrm{Fe}\right]$ of the MW knot in bins weighted by the GMM probabilities, to compare the observational measurements with our results. 

Figure~\ref{fig:horta} shows the median iron and alpha-element abundances, traced by Mg (MW observations; yellow star) and by O for the simulated galaxies (this work; circles). O and Mg share the same nucleosynthetic origin but are not strictly equivalent. \textsc{CIELO} only tracks $^{24}$Mg (see Section~\ref{sec:CIELO}), resulting in systematically underestimated $\left[\mathrm{Mg}/\mathrm{Fe}\right]$ values. We therefore adopt O as our alpha-element tracer. In this comparison, it is also important to keep in mind that observed stars were classified according to their GMM membership probability, rather than to a strict separation among components. Median points in the $\left[\alpha/\mathrm{Fe}\right]$--$\left[\mathrm{Fe}/\mathrm{H}\right]$ plane constrain enrichment timescales and can be used as chemical clocks (see Section~\ref{sec:charact}). The observed MW knot abundances fall within the range of \textsc{CIELO} knots, close to the two knots hosted by MW-mass-range galaxies, with a slight offset toward lower $\left[\mathrm{Fe}/\mathrm{H}\right]$.

We find a correlation between the overall median $\left[\mathrm{O}/\mathrm{Fe}\right]$ and $\left[\mathrm{Fe}/\mathrm{H}\right]$ values ($r=0.52$, $p<10^{-6}$), implying that bulges with higher iron abundances also show higher alpha-element abundances. This seems at odds with \citet{iza+2025}, who found anti-correlations for Auriga bulge, halo, and disk components; however, that work analyzed only MW-mass galaxies, where anti-correlations at fixed mass are expected from star formation history modulation~\citep{Tinsley1979}. Our sample instead spans a wider range of galaxy stellar masses, and Fig.~\ref{fig:horta} shows that galaxies of similar mass do follow anti-correlation trends, shifting to lower $\left[\mathrm{Fe}/\mathrm{H}\right]$ at lower $\left[\mathrm{O}/\mathrm{Fe}\right]$~\citep{matteucci+brocato1990}, as expected from the mass-metallicity relation.

\citet{gonzalez-jara+2025} found a similar correlation for the \textsc{CIELO} stellar halo, attributed to low-mass accreted satellites with more continuous star formation histories, allowing SNIa to enrich the ISM and decrease its alpha-to-iron abundance ratio. We acknowledge that our sample is small for a more detailed statistical analysis; the $\left[\alpha/\mathrm{Fe}\right]$ as a function of $\left[\mathrm{Fe}/\mathrm{H}\right]$ will be further addressed in Gonzalez-Jara et al. (in prep.) in a global context.

\subsection{Comparison with TNG50}
\begin{figure}[ht!]
   \centering
   \includegraphics[width=1\hsize]{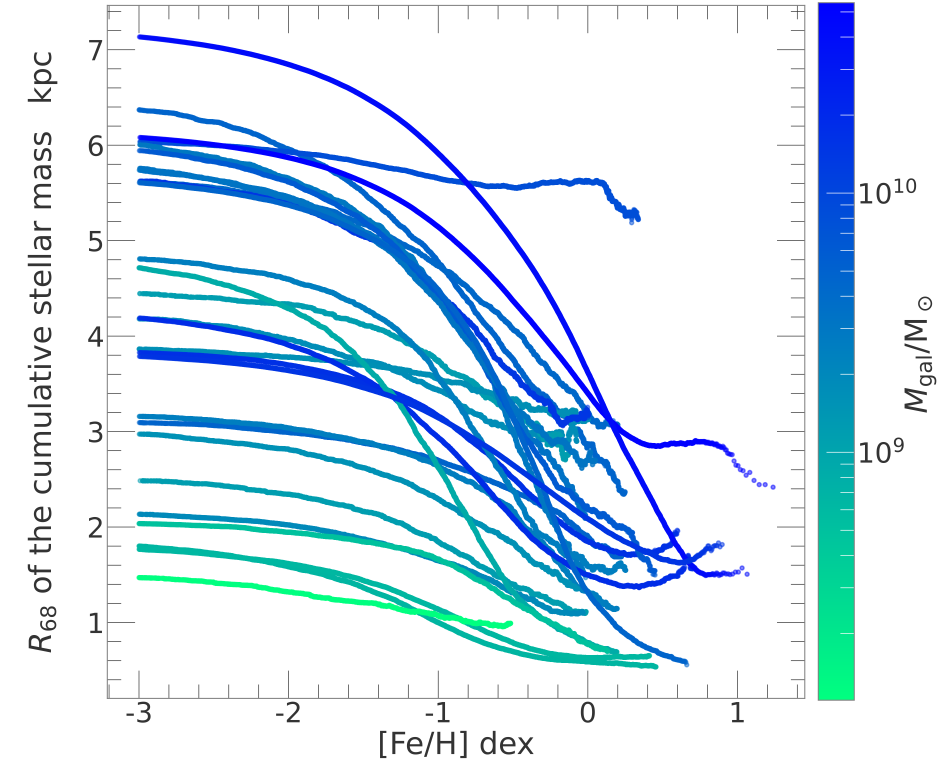}
      \caption{Cumulative radial extent of metal-rich stellar populations for knotted \textsc{CIELO} galaxies. For each $\left[\mathrm{Fe}/\mathrm{H}\right]$ threshold, $R_{68}$ is the radius enclosing $68\%$ of stellar mass with iron abundance equal to or greater than that value. Curves are color-coded by galaxy stellar mass.}
         \label{fig:rix}
\end{figure}
\citet{rix+2024_mrichknot} focused on studying the spatial distribution of the MW's most metal-rich stars ($\left[\mathrm{M}/\mathrm{H}\right]\gtrsim 0.5$) using Gaia XP spectroscopy, finding these stars tightly confined within the inner $1.5\,$kpc. Although they applied a different selection approach, they argued that their sample overlaps with that of \citet{horta+2025_knot}, thus adopting the same ``knot'' terminology. To provide theoretical context, they also compared their results with MW-like galaxies from the TNG50 cosmological simulations to determine the expected spatial extent of extremely metal-rich stars.

To compare \textsc{CIELO} with the results from TNG50, we employed a similar approach to that \citet{rix+2024_mrichknot} with our data from \textsc{CIELO}. In our case, we used all stellar particles from the 28 knotted galaxies and performed a cumulative analysis from high to low iron abundance, determining at each iron abundance threshold the radial extent ($R_{68}$) that encloses $68\%$ of stellar mass with iron abundances equal to or greater than that threshold. In other words, for each limiting iron abundance value, we accumulate stellar mass from the metal-rich end and measure how spatially confined this cumulative population is. This approach captures the extent of progressively more metal-rich stellar populations, and it is not a mono-abundance approach. This procedure generates the $R_{68}$ versus iron abundance relationships shown in Fig.~\ref{fig:rix}.

Since the \textsc{CIELO} sample spans a wide range of galaxy masses, both the extent of the most metal-rich populations and the maximum iron abundance reached vary accordingly. Remarkably, the majority of the \textsc{CIELO} knotted galaxies demonstrate that metal-rich populations are rather highly localized, with metals centrally confined to compact knots of $\lesssim3\,$kpc. According to our selection of 28 knots, they have sizes ($r_{90}$) between $0.33$--$1.39\,$kpc\footnote{The two smallest objects (P7-0192 with $r_{90} = 0.33\,$kpc and P7-2780 with $r_{90} =0.41\,$kpc) correspond to the higher resolution run, with sizes $>\epsilon_\mathrm{grav}$.}. This result is in agreement with that of \citet{rix+2024_mrichknot}, in which TNG50 MW-mass galaxies have their most metal-rich stars centrally concentrated in regions of $\lesssim 1\,$kpc, although in our case the transition toward higher iron abundance at small radii is less abrupt and more gradual. In the case of TNG50, recent results have shown their simulated galaxies to have high iron abundance, dense star-forming regions which might be the consequence of specific feedback implementation \citep{celiz+2025, hu+2026}. This partial agreement is encouraging, given that \textsc{CIELO} and TNG50 differ in their numerical approach (SPH vs. moving-mesh), feedback prescriptions, and chemical evolution modeling \citep{naiman+2018}; as such, it should not be regarded as validating either model, but rather as motivation to further investigate these central overdensities with independent numerical approaches.

\subsection{Enrichment prescriptions: caveats}
The interpretation of small differences in $\left[\mathrm{O}/\mathrm{Fe}\right]$ and $\left[\mathrm{Fe}/\mathrm{H}\right]$ as evidence for distinct formation pathways relies on the enrichment prescriptions adopted in the \textsc{CIELO} chemical evolution model, including the IMF, massive-star yields and explosion prescriptions, the SNIa delay-time distribution, feedback-ISM energy coupling, and metal mixing/partition between gas phases. These choices suit the broader goals of galaxy formation studies, but quantitative abundance differences of a few hundredths of a dex, such as those presented in Section~\ref{sec:charact}, may be sensitive to them, an issue currently being explored by testing alternative SNIa delay-time distributions (Jara-Ferreira et al., in prep.); \citet{yan+2020}, for instance, show that IMF assumptions can significantly affect predicted chemical evolution in galaxies forming on short timescales, particularly relevant here given that knots form on timescales comparable to the SNIa delay time. Improved stellar population sampling, such as the semi-deterministic approach of \citet{deng+2026}, can further affect predicted enrichment histories in low-mass galaxies, where stochastic sampling of massive stars matters. The fact that our results are based on a relative comparison between knots, bulge, and co-spatial populations certainly reduces, although does not necessarily erase, the impact of the enrichment prescriptions. These caveats do not affect the qualitative trends reported throughout this work, but should be kept in mind when interpreting the specific quantitative values discussed, including the comparisons with independent works presented above.

\section{Conclusions}
\label{sec:conc}
We have studied knot structures at the centers of simulated galaxies from the \textsc{CIELO} project~\citep{tissera+2025}, with stellar masses in the range $10^{8.0}$--$10^{10.7}\,\mathrm{M}_\odot$. Our methodology identifies these structures as overdensities in the energy and circularity distributions in $(\epsilon, E)$ space, focusing on the most gravitationally bound stellar populations at galactic centers. After applying selection criteria based on energy peak identification, compactness relative to the bulge ($r_{90}/r_{50}$ ratio), and morphological coherence, we isolated a sample of 28 knots in the centers of galaxies drawn from a parent sample of 54 galaxies. We carried out a numerical resolution study finding that in intermediate numerical resolution runs of \textsc{CIELO} the number of knots might be underestimated. Therefore, here we do not focus on the frequency of knots in galaxies, only in the relative comparison of the properties of the stellar populations located in bulge resolved with similar numerical resolution. Nonetheless, resolution sensitivity persists and is expected to propagate to the physical properties characterized throughout this work (see Appendix~\ref{sec:appendix_reseff}). Given this, our results should be regarded as plausible trends in central stellar knots, rather than a definitive characterization of their physical properties.

By comparing these dynamically decoupled knots to their reference co-spatial and bulge components (excluding knot particles from both), we identified significant chemical differences revealing distinct evolutionary pathways. Our main finding is an enhancement in $\left[\mathrm{O}/\mathrm{Fe}\right]$ ratios for the knots (Figs.~\ref{fig:alphaplane_median} and \ref{fig:alphaplane_mode}), consistent with the shorter formation timescales we also find for these structures (Fig.~\ref{fig:t5090}). All three components share similar in-situ/accreted distributions (Fig.~\ref{fig:acc_disk_boxplots}a), though knots have a slightly lower median accreted fraction ($5\%$ vs. $\sim$10$\%$). Disk-born fractions are negligible for knots ($\sim$0$\%$) compared to other components ($\sim$10$\%$; Fig.~\ref{fig:acc_disk_boxplots}b), except in massive MW-like systems ($10$--$20\%$; Fig.~\ref{fig:accreted_disk_individual}b). Gas tracking (Fig.~\ref{fig:gas_origin_boxplots}) shows knots receive their largest contribution from external infall, followed by fossil bulge gas, then a primordial disk; fossil bulge gas is consistent with being the primary fuel for their iron and alpha enrichment (Fig.~\ref{fig:stellar_gaschannel_FeHOFe_boxplots}), with knots showing the largest fossil bulge fraction among all components ($21\%$ vs. $17$--$16\%$), consistent with early-formed, centrally confined structures. Finally, knots exhibit various morphologies (prolate, triaxial, oblate) but are primarily spheroidal ($54\%$; Fig.~\ref{fig:triax}), distributed in roughly symmetric structures deep within the potential wells, which are more efficient at retaining metals.

When compared with the literature, \textsc{CIELO} knots show good agreement with both observational and simulation-based studies of the MW's central stellar populations. In the $\left[\alpha/\mathrm{Fe}\right]$ -- $\left[\mathrm{Fe}/\mathrm{H}\right]$ plane, the MW knot identified by \citet{horta+2025_knot} from APOGEE--Gaia data lies within the region populated by \textsc{CIELO} knots, which systematically show higher iron and alpha-element abundances than their co-spatial and bulge counterparts. Additionally, following the approach of \citet{rix+2024_mrichknot}, we found that the most metal-rich stellar populations in \textsc{CIELO} knotted galaxies are highly spatially confined, with sizes of $3\,$kpc, consistent with the centrally concentrated metal-rich structures found in TNG50 MW-mass galaxies. Our knots themselves have sizes ($r_{90}$) ranging from $0.33$ to $1.39$\,kpc, consistently falling within this metal-rich central region across our broader range of host galaxy stellar masses. Together, these comparisons place our work in a broader context, bridging an independent simulation framework with observational and other simulation-based constraints on the nature and existence of stellar knots residing at the bottom of galactic potential wells.

We are aware that our findings would benefit from a larger statistical sample and higher numerical resolution. Given the resolution sensitivity discussed above, we distinguish between our most robust and more tentative findings. Among the former, the consistency across independent diagnostics, namely alpha-element enhancement, formation timescales, and accreted/disk-born fractions, constitutes a coherent formation scenario. This is further supported by the markedly different properties of co-spatial stars, which validates our dynamical selection in $(\epsilon,E)$ space and is of broader relevance for dynamical decontamination in overlapping galactic regions. On the other side, knot occurrence frequency and the exact structural and chemical abundance values reported here should be considered more tentative, as they depend on precisely isolating a knot in the midst of its host galaxy, which remains sensitive to numerical resolution.

The origin of the $\sigma$-peak~\citep{zoccali+2014, valenti+2018, quezada2025} and the potential contribution of structures such as stellar knots to this kinematic feature remain open questions. To address this, we are currently comparing the knots identified here with observational data from the updated MW Bulge kinematic map of \citet{quezada2025}, projecting our results onto the observational space (Acosta-Tripailao et al., in prep.).

Knots may hold particular relevance beyond bulge assembly itself. It is worth noting a possible connection between stellar knots and the early growth of supermassive black holes (SMBHs). Knots mark the densest and most gravitationally bound regions of their host galaxies, characterized by rapid, early, in-situ star formation, precisely the kind of environment favored by SMBHs for efficient gas accretion in the early Universe~\citep{shi+2024}. Although \textsc{CIELO} does not model SMBHs, this potential connection motivates future work exploring whether stellar knots trace the preferential sites for early SMBH growth.

\begin{acknowledgements}
We thank D. Horta for providing the data used in Fig.~\ref{fig:horta}. BAT, JGJ, and BTC acknowledge funding by ANID (Beca Doctorado Nacional, Folios 21231305, 21210846, and 21232155, respectively). IME acknowledges funding by ANID (Beca Magíster Nacional, Folio 22241737). PBT and JGJ acknowledge partial funding by ANID BASAL project FB210003; PBT also acknowledges FONDECYT-ANID 1240465/2024. MZ acknowledges FONDECYT REGULAR grant No. 1230731. This project has received funding from the European Union Horizon 2020 Research and Innovation Programme under Marie Skłodowska-Curie Actions (MSCA) grant agreement No. 101086388-LACEGAL, and used the Ladgerda Cluster (FONDECYT 1200703/2020, Institute for Astrophysics, Chile), Marenostrum 4 (Barcelona SuperComputer Center, Spain), the NLHPC (Centro de Modelamiento Matem\'atico, Chile), and Geryon clusters (Center for Astrophysics, CATA, Chile).
\end{acknowledgements}

\bibliographystyle{aa}
\bibliography{biblio}

\begin{thebibliography}{85}
\expandafter\ifx\csname natexlab\endcsname\relax\def\natexlab#1{#1}\fi

\bibitem[{{Aguerri} {et~al.}(2001){Aguerri}, {Balcells}, \&
  {Peletier}}]{aguerri+2001}
{Aguerri}, J.~A.~L., {Balcells}, M., \& {Peletier}, R.~F. 2001, \aap, 367, 428

\bibitem[{{Babusiaux}(2016)}]{babusiaux2016}
{Babusiaux}, C. 2016, \pasa, 33, e026

\bibitem[{{Babusiaux} {et~al.}(2014){Babusiaux}, {Katz}, {Hill}, {Royer},
  {G{\'o}mez}, {Arenou}, {Combes}, {Di Matteo}, {Gilmore}, {Haywood}, {Robin},
  {Rodriguez-Fernandez}, {Sartoretti}, \& {Schultheis}}]{babusiaux+2014}
{Babusiaux}, C., {Katz}, D., {Hill}, V., {et~al.} 2014, \aap, 563, A15

\bibitem[{{Baker} {et~al.}(2023){Baker}, {Maiolino}, {Belfiore}, {Curti},
  {Bluck}, {Lin}, {Ellison}, {Thorp}, \& {Pan}}]{baker2023}
{Baker}, W.~M., {Maiolino}, R., {Belfiore}, F., {et~al.} 2023, \mnras, 519,
  1149

\bibitem[{{Casanueva} {et~al.}(2022){Casanueva}, {Lagos}, {Padilla}, \&
  {Davison}}]{Casanueva+2022}
{Casanueva}, C.~I., {Lagos}, C. d.~P., {Padilla}, N.~D., \& {Davison}, T.~A.
  2022, \mnras, 514, 2031

\bibitem[{{Cataldi} {et~al.}(2023){Cataldi}, {Pedrosa}, {Tissera}, {Artale},
  {Padilla}, {Dominguez-Tenreiro}, {Bignone}, {Gonzalez}, \&
  {Pellizza}}]{cataldi_2023}
{Cataldi}, P., {Pedrosa}, S.~E., {Tissera}, P.~B., {et~al.} 2023, \mnras, 523,
  1919

\bibitem[{{Celiz} {et~al.}(2025){Celiz}, {Navarro}, {Abadi}, \&
  {Springel}}]{celiz+2025}
{Celiz}, B.~M., {Navarro}, J.~F., {Abadi}, M.~G., \& {Springel}, V. 2025, \aap,
  699, A12

\bibitem[{Chabrier(2003)}]{chabrier2003}
Chabrier, G. 2003, \apj, 586, L133–L136

\bibitem[{Cormen {et~al.}(2001)Cormen, Leiserson, Rivest, \&
  Stein}]{Cormen2001}
Cormen, T.~H., Leiserson, C.~E., Rivest, R.~L., \& Stein, C. 2001, Introduction
  to Algorithms, 2nd edn. (Cambridge, MA: MIT Press)

\bibitem[{{Costantin} {et~al.}(2018){Costantin}, {M{\'e}ndez-Abreu}, {Corsini},
  {Eliche-Moral}, {Tapia}, {Morelli}, {Dalla Bont{\`a}}, \&
  {Pizzella}}]{costantin+2018b}
{Costantin}, L., {M{\'e}ndez-Abreu}, J., {Corsini}, E.~M., {et~al.} 2018, \aap,
  609, A132

\bibitem[{{Davis} {et~al.}(1985){Davis}, {Efstathiou}, {Frenk}, \&
  {White}}]{davis+1985}
{Davis}, M., {Efstathiou}, G., {Frenk}, C.~S., \& {White}, S.~D.~M. 1985, \apj,
  292, 371

\bibitem[{{De Leo} {et~al.}(2026){De Leo}, {Massari}, {Bellazzini},
  {Mucciarelli}, {Acosta-Tripailao}, \& {Nipoti}}]{deleo+2025_dynamicalmirages}
{De Leo}, M., {Massari}, D., {Bellazzini}, M., {et~al.} 2026, \aap, 707, A310

\bibitem[{{Dekel} \& {Silk}(1986)}]{dekel+silk1986}
{Dekel}, A. \& {Silk}, J. 1986, \apj, 303, 39

\bibitem[{{Deng} {et~al.}(2026){Deng}, {Li}, {Yan}, {Kong}, \&
  {Zhang}}]{deng+2026}
{Deng}, Y., {Li}, H., {Yan}, Z., {Kong}, C., \& {Zhang}, Z.-Y. 2026, ArXiv
  e-prints [\eprint[arXiv]{2606.09999}], submitted to AAS Journals

\bibitem[{Dillamore \& Sanders(2025)}]{dillamore2025}
Dillamore, A.~M. \& Sanders, J.~L. 2025, \mnras, 542, 1331

\bibitem[{{Dolag} {et~al.}(2009){Dolag}, {Borgani}, {Murante}, \&
  {Springel}}]{dolag+2009}
{Dolag}, K., {Borgani}, S., {Murante}, G., \& {Springel}, V. 2009, \mnras, 399,
  497

\bibitem[{{Edgeworth}(1888)}]{Edgeworth1888}
{Edgeworth}, F.~Y. 1888, Philos. Mag., 25, 184

\bibitem[{{Ferrarese} \& {Merritt}(2000)}]{ferrarese+merrit2000}
{Ferrarese}, L. \& {Merritt}, D. 2000, \apjl, 539, L9

\bibitem[{{Fragkoudi} {et~al.}(2020){Fragkoudi}, {Grand}, {Pakmor},
  {Bl{\'a}zquez-Calero}, {Gargiulo}, {Gomez}, {Marinacci}, {Monachesi}, {Ness},
  {Perez}, {Tissera}, \& {White}}]{fragkoudi+2020_mwquiescent}
{Fragkoudi}, F., {Grand}, R.~J.~J., {Pakmor}, R., {et~al.} 2020, \mnras, 494,
  5936

\bibitem[{{Fragkoudi} {et~al.}(2025){Fragkoudi}, {Grand}, {Pakmor},
  {G{\'o}mez}, {Marinacci}, \& {Springel}}]{fragkoudi+2025}
{Fragkoudi}, F., {Grand}, R. J.~J., {Pakmor}, R., {et~al.} 2025, \mnras, 538,
  1587

\bibitem[{{Gallazzi} {et~al.}(2005){Gallazzi}, {Charlot}, {Brinchmann},
  {White}, \& {Tremonti}}]{Gallazzi2005}
{Gallazzi}, A., {Charlot}, S., {Brinchmann}, J., {White}, S. D.~M., \&
  {Tremonti}, C.~A. 2005, \mnras, 362, 41

\bibitem[{{Gargiulo} {et~al.}(2019){Gargiulo}, {Monachesi}, {G{\'o}mez},
  {Grand}, {Marinacci}, {Pakmor}, {White}, {Bell}, {Fragkoudi}, \&
  {Tissera}}]{gargiulo+2019}
{Gargiulo}, I.~D., {Monachesi}, A., {G{\'o}mez}, F.~A., {et~al.} 2019, \mnras,
  489, 5742

\bibitem[{{Gonzalez-Jara} {et~al.}(2025){Gonzalez-Jara}, B., {Monachesi},
  {Sillero}, {Pallero}, {Pedrosa}, {Tau}, {Tapia-Contreras}, \&
  {Bignone}}]{gonzalez-jara+2025}
{Gonzalez-Jara}, J., B., T., {Monachesi}, A., {et~al.} 2025, \aap, 693, A282

\bibitem[{{Grand} {et~al.}(2017){Grand}, {G{\'o}mez}, {Marinacci}, {Pakmor},
  {Springel}, {Campbell}, {Frenk}, {Jenkins}, \& {White}}]{grand+2017_auriga}
{Grand}, R. J.~J., {G{\'o}mez}, F.~A., {Marinacci}, F., {et~al.} 2017, \mnras,
  467, 179

\bibitem[{{Gudin} {et~al.}(2021){Gudin}, {Shank}, {Beers}, {Yuan}, {Limberg},
  {Roederer}, {Placco}, {Holmbeck}, {Dietz}, {Rasmussen}, {Hansen}, {Sakari},
  {Ezzeddine}, \& {Frebel}}]{gudin2021}
{Gudin}, D., {Shank}, D., {Beers}, T.~C., {et~al.} 2021, \apj, 908, 79

\bibitem[{{G{\"u}ltekin} {et~al.}(2009){G{\"u}ltekin}, {Richstone}, {Gebhardt},
  {Lauer}, {Tremaine}, {Aller}, {Bender}, {Dressler}, {Faber}, {Filippenko},
  {Green}, {Ho}, {Kormendy}, {Magorrian}, {Pinkney}, \&
  {Siopis}}]{gultekin+2009}
{G{\"u}ltekin}, K., {Richstone}, D.~O., {Gebhardt}, K., {et~al.} 2009, \apj,
  698, 198

\bibitem[{Hahn \& Abel(2011)}]{hahn&abel2011}
Hahn, O. \& Abel, T. 2011, \mnras, 415, 2101

\bibitem[{{Hasselquist} {et~al.}(2020){Hasselquist}, {Zasowski}, {Feuillet},
  {Schultheis}, {Nataf}, {Anguiano}, {Beaton}, {Beers}, {Cohen}, {Cunha},
  {Fern{\'a}ndez-Trincado}, {Garc{\'\i}a-Hern{\'a}ndez}, {Geisler}, {Holtzman},
  {Johnson}, {Lane}, {Majewski}, {Moni Bidin}, {Nitschelm}, {Roman-Lopes},
  {Schiavon}, {Smith}, \& {Sobeck}}]{hasselquist2020}
{Hasselquist}, S., {Zasowski}, G., {Feuillet}, D.~K., {et~al.} 2020, \apj, 901,
  109

\bibitem[{{Horta Darrington} {et~al.}(2025){Horta Darrington}, {Petersen}, \&
  {Pe{\~n}arrubia}}]{horta+2025_knot}
{Horta Darrington}, D., {Petersen}, M.~S., \& {Pe{\~n}arrubia}, J. 2025,
  \mnras, 538, 998

\bibitem[{Hu {et~al.}(2026)Hu, Yang, \& Gao}]{hu+2026}
Hu, J., Yang, H., \& Gao, L. 2026, \apj, 1008, 54

\bibitem[{Iwamoto {et~al.}(1999)Iwamoto, Brachwitz, Nomoto, Kishimoto, Umeda,
  Hix, \& Thielemann}]{iwamoto+1999}
Iwamoto, K., Brachwitz, F., Nomoto, K., {et~al.} 1999, \apjs, 125, 439

\bibitem[{{Iza} {et~al.}(2025){Iza}, {Scannapieco}, {Nuza}, {Pakmor}, {Grand},
  {Gómez}, {Springel}, {Marinacci}, \& {Fragkoudi}}]{iza+2025}
{Iza}, F.~G., {Scannapieco}, C., {Nuza}, S.~E., {et~al.} 2025, \aap, 701, A99

\bibitem[{Jackson {et~al.}(2021)Jackson, Jofré, Yaxley, Das, de~Brito~Silva,
  \& Foley}]{jackson2021}
Jackson, H., Jofré, P., Yaxley, K., {et~al.} 2021, \mnras, 502, 32

\bibitem[{Jiménez {et~al.}(2015)Jiménez, Tissera, \&
  Matteucci}]{jimenez+2015}
Jiménez, N., Tissera, P.~B., \& Matteucci, F. 2015, \apj, 810, 137

\bibitem[{Jofré {et~al.}(2017)Jofré, Das, Bertranpetit, \&
  Foley}]{jofre2017_phylogeny}
Jofré, P., Das, P., Bertranpetit, J., \& Foley, R. 2017, \mnras, 467, 1140

\bibitem[{{Joyce} {et~al.}(2023){Joyce}, {Johnson}, {Marchetti}, {Rich},
  {Simion}, \& {Bourke}}]{joyce2023}
{Joyce}, M., {Johnson}, C.~I., {Marchetti}, T., {et~al.} 2023, \apj, 946, 28

\bibitem[{Keith {et~al.}(2025)Keith, Munshi, Brooks, Van~Nest, Engelhardt,
  Cruz, Keller, Quinn, \& Wadsley}]{Keith_2025}
Keith, B., Munshi, F., Brooks, A.~M., {et~al.} 2025, \apj, 986, 138

\bibitem[{Knollmann \& Knebe(2009)}]{Knollmann_2009}
Knollmann, S.~R. \& Knebe, A. 2009, \apjs, 182, 608

\bibitem[{{Kormendy} \& {Ho}(2013)}]{kormendy+ho2013}
{Kormendy}, J. \& {Ho}, L.~C. 2013, \araa, 51, 511

\bibitem[{Lian {et~al.}(2025)Lian, Wang, Feng, Huang, \& Guo}]{lian_2025}
Lian, J., Wang, T., Feng, Q., Huang, Y., \& Guo, H. 2025, \apjl, 990, L37

\bibitem[{López {et~al.}(2025)López, Fragkoudi, Cora, Scannapieco, Pakmor,
  Grand, Gómez, \& Marinacci}]{lopez+2025}
López, P.~D., Fragkoudi, F., Cora, S.~A., {et~al.} 2025, \mnras, 540, 2031

\bibitem[{{Martig} {et~al.}(2021){Martig}, {Pinna}, {Falc{\'o}n-Barroso},
  {Gadotti}, {Husemann}, {Minchev}, {Neumann}, {Ruiz-Lara}, \& {van de
  Ven}}]{martig+2021}
{Martig}, M., {Pinna}, F., {Falc{\'o}n-Barroso}, J., {et~al.} 2021, \mnras,
  508, 2458

\bibitem[{{Matteucci} \& {Brocato}(1990)}]{matteucci+brocato1990}
{Matteucci}, F. \& {Brocato}, E. 1990, \apj, 365, 539

\bibitem[{{McWilliam}(1997)}]{McWilliam1997}
{McWilliam}, A. 1997, \araa, 35, 503

\bibitem[{Mosconi {et~al.}(2001)Mosconi, Tissera, Lambas, \&
  Cora}]{mosconi+2001}
Mosconi, M.~B., Tissera, P.~B., Lambas, D.~G., \& Cora, S.~A. 2001, \mnras,
  325, 34

\bibitem[{{Mu{\~n}oz-Escobar} {et~al.}(2026){Mu{\~n}oz-Escobar}, {Tissera},
  {Gonzalez-Jara}, {Sillero}, {Miranda}, {Pedrosa}, \&
  {Bignone}}]{munoz-escobar2026}
{Mu{\~n}oz-Escobar}, I., {Tissera}, P.~B., {Gonzalez-Jara}, J., {et~al.} 2026,
  \aap, 705, A87

\bibitem[{Naiman {et~al.}(2018)Naiman, Pillepich, Springel, Ramirez-Ruiz,
  Torrey, Vogelsberger, Pakmor, Nelson, Marinacci, Hernquist, Weinberger, \&
  Genel}]{naiman+2018}
Naiman, J.~P., Pillepich, A., Springel, V., {et~al.} 2018, \mnras, 477, 1206

\bibitem[{Pillepich {et~al.}(2024)Pillepich, Sotillo-Ramos, Ramesh, Nelson,
  Engler, Rodriguez-Gomez, Fournier, Donnari, Springel, \&
  Hernquist}]{pillepich_2024}
Pillepich, A., Sotillo-Ramos, D., Ramesh, R., {et~al.} 2024, \mnras, 535, 1721

\bibitem[{{Quezada} {et~al.}(2025){Quezada}, {Zoccali}, {Valenti}, {Rojas
  Arriagada}, {Renzini}, {Gonzalez}, {Mucciarelli}, {Rejkuba}, {Surot}, \&
  {Valenzuela Navarro}}]{quezada2025}
{Quezada}, C., {Zoccali}, M., {Valenti}, E., {et~al.} 2025, \aap, 702, A164

\bibitem[{{Renzini} {et~al.}(2018){Renzini}, {Gennaro}, {Zoccali}, {Brown},
  {Anderson}, {Minniti}, {Sahu}, {Valenti}, \& {VandenBerg}}]{renzini2018}
{Renzini}, A., {Gennaro}, M., {Zoccali}, M., {et~al.} 2018, \apj, 863, 16

\bibitem[{{Rix} {et~al.}(2024){Rix}, {Chandra}, {Zasowski}, {Pillepich},
  {Khoperskov}, {Feltzing}, {Wyse}, {Frankel}, {Horta}, {Kollmeier}, {Stassun},
  {Ness}, {Bird}, {Nidever}, {Fern{\'a}ndez-Trincado}, {Amarante}, {Laporte},
  \& {Lian}}]{rix+2024_mrichknot}
{Rix}, H.-W., {Chandra}, V., {Zasowski}, G., {et~al.} 2024, \apj, 975, 293

\bibitem[{{Romano} {et~al.}(2023){Romano}, {Ferraro}, {Origlia}, {Portegies
  Zwart}, {Lanzoni}, {Crociati}, {Massari}, {Dalessandro}, {Mucciarelli},
  {Rich}, {Calura}, \& {Matteucci}}]{romano+2023}
{Romano}, D., {Ferraro}, F.~R., {Origlia}, L., {et~al.} 2023, \apj, 951, 85

\bibitem[{{Savino} {et~al.}(2020){Savino}, {Koch}, {Prudil}, {Kunder}, \&
  {Smolec}}]{savino2020}
{Savino}, A., {Koch}, A., {Prudil}, Z., {Kunder}, A., \& {Smolec}, R. 2020,
  \aap, 641, A96

\bibitem[{{Scannapieco} {et~al.}(2005){Scannapieco}, {Tissera}, {White}, \&
  {Springel}}]{scannapieco+2005}
{Scannapieco}, C., {Tissera}, P.~B., {White}, S.~D.~M., \& {Springel}, V. 2005,
  \mnras, 364, 552

\bibitem[{Scannapieco {et~al.}(2006)Scannapieco, Tissera, White, \&
  Springel}]{scannapieco+2006}
Scannapieco, C., Tissera, P.~B., White, S. D.~M., \& Springel, V. 2006, \mnras,
  371, 1125

\bibitem[{{Scannapieco} {et~al.}(2008){Scannapieco}, {Tissera}, {White}, \&
  {Springel}}]{scannapieco2008}
{Scannapieco}, C., {Tissera}, P.~B., {White}, S. D.~M., \& {Springel}, V. 2008,
  \mnras, 389, 1137

\bibitem[{Shi {et~al.}(2024)Shi, Kremer, \& Hopkins}]{shi+2024}
Shi, Y., Kremer, K., \& Hopkins, P.~F. 2024, \apjl, 969, L31

\bibitem[{{Somerville} \& {Dav{\'e}}(2015)}]{somerville_dave2015}
{Somerville}, R.~S. \& {Dav{\'e}}, R. 2015, \araa, 53, 51

\bibitem[{{Speagle} {et~al.}(2014){Speagle}, {Steinhardt}, {Capak}, \&
  {Silverman}}]{speagle2014}
{Speagle}, J.~S., {Steinhardt}, C.~L., {Capak}, P.~L., \& {Silverman}, J.~D.
  2014, \apjs, 214, 15

\bibitem[{{Springel} {et~al.}(2005){Springel}, {Di Matteo}, \&
  {Hernquist}}]{springel2005}
{Springel}, V., {Di Matteo}, T., \& {Hernquist}, L. 2005, \apjl, 620, L79

\bibitem[{{Springel} \& {Hernquist}(2003)}]{springel2003}
{Springel}, V. \& {Hernquist}, L. 2003, \mnras, 339, 312

\bibitem[{{Springel} {et~al.}(2001){Springel}, {White}, {Tormen}, \&
  {Kauffmann}}]{springel+2001}
{Springel}, V., {White}, S. D.~M., {Tormen}, G., \& {Kauffmann}, G. 2001,
  \mnras, 328, 726

\bibitem[{Starkenburg {et~al.}(2017)Starkenburg, Martin, Youakim, Aguado,
  Prieto, Arentsen, Bernard, Bonifacio, Caffau, Carlberg, C{\^o}t{\'e},
  Fouesneau, Fran{\c{c}}ois, Franke, Hern{\'a}ndez, Gwyn, Hill, Ibata,
  Jablonka, Longeard, McConnachie, Navarro, S{\'a}nchez-Janssen, Tolstoy, \&
  Venn}]{starkenburg+2017}
Starkenburg, E., Martin, N.~F., Youakim, K., {et~al.} 2017, \mnras, 471, 2587

\bibitem[{{Tapia-Contreras} {et~al.}(2025){Tapia-Contreras}, {Tissera},
  {Sillero}, {Gonzalez-Jara}, {Casanueva-Villarreal}, {Pedrosa}, {Bignone},
  {Padilla}, \& {Dom{\'\i}nguez-Tenreiro}}]{tapia-contreras+2025}
{Tapia-Contreras}, B., {Tissera}, P.~B., {Sillero}, E., {et~al.} 2025, \aap,
  700, A69

\bibitem[{{Tapia-Contreras} {et~al.}(2026){Tapia-Contreras}, {Tissera},
  {Sillero}, {Jofr{\'e}}, {Yaxley}, {Hua}, {Yates}, {M{\'a}rquez S.}, {Signor},
  {Das}, {Rojas-Arriagada}, {Aguilera-G{\'o}mez}, {Jara-Ferreira}, \&
  {Foley}}]{tapia-contreras+2026}
{Tapia-Contreras}, B., {Tissera}, P.~B., {Sillero}, E., {et~al.} 2026, \aap,
  711, A263

\bibitem[{{Thob} {et~al.}(2019){Thob}, {Crain}, {McCarthy}, {Schaller},
  {Lagos}, {Schaye}, {Talens}, {James}, {Theuns}, \& {Bower}}]{Thob+19}
{Thob}, A. C.~R., {Crain}, R.~A., {McCarthy}, I.~G., {et~al.} 2019, \mnras,
  485, 972

\bibitem[{{Tinsley}(1979)}]{Tinsley1979}
{Tinsley}, B.~M. 1979, \apj, 229, 1046

\bibitem[{{Tissera} {et~al.}(2025){Tissera}, {Bignone}, {Gonzalez-Jara},
  {Mu{\~n}oz-Escobar}, {Cataldi}, {Miranda}, {Barrientos-Acevedo},
  {Tapia-Contreras}, {Pedrosa}, {Padilla}, {Dominguez-Tenreiro},
  {Casanueva-Villarreal}, {Sillero}, {Silva-Mella}, {Shailesh}, \&
  {Jara-Ferreira}}]{tissera+2025}
{Tissera}, P.~B., {Bignone}, L., {Gonzalez-Jara}, J., {et~al.} 2025, \aap, 697,
  A134

\bibitem[{{Tissera} \& {Dominguez-Tenreiro}(1998)}]{tissera_1998}
{Tissera}, P.~B. \& {Dominguez-Tenreiro}, R. 1998, \mnras, 297, 177

\bibitem[{Tissera {et~al.}(2017)Tissera, Machado, Carollo, Minniti, Beers,
  Zoccali, \& Meza}]{tissera+zoccali+2017}
Tissera, P.~B., Machado, R. E.~G., Carollo, D., {et~al.} 2017, \mnras, 473,
  1656

\bibitem[{Tissera {et~al.}(2011)Tissera, White, \& Scannapieco}]{tissera+2011}
Tissera, P.~B., White, S. D.~M., \& Scannapieco, C. 2011, \mnras, 420,
  255–270

\bibitem[{{Tissera} {et~al.}(2012){Tissera}, {White}, \&
  {Scannapieco}}]{tissera2012}
{Tissera}, P.~B., {White}, S. D.~M., \& {Scannapieco}, C. 2012, \mnras, 420,
  255

\bibitem[{{Tremonti} {et~al.}(2004){Tremonti}, {Heckman}, {Kauffmann},
  {Brinchmann}, {Charlot}, {White}, {Seibert}, {Peng}, {Schlegel}, {Uomoto},
  {Fukugita}, \& {Brinkmann}}]{tremonti2004}
{Tremonti}, C.~A., {Heckman}, T.~M., {Kauffmann}, G., {et~al.} 2004, \apj, 613,
  898

\bibitem[{{Valenti} {et~al.}(2018){Valenti}, {Zoccali}, {Mucciarelli},
  {Gonzalez}, {Surot}, {Minniti}, \& {Rejkuba}}]{valenti+2018}
{Valenti}, E., {Zoccali}, M., {Mucciarelli}, A., {et~al.} 2018, \aap, 616, A83

\bibitem[{{Valenzuela} {et~al.}(2024){Valenzuela}, {Remus}, {Dolag}, \&
  {Seidel}}]{valenzuela_remus2024}
{Valenzuela}, L.~M., {Remus}, R.-S., {Dolag}, K., \& {Seidel}, B.~A. 2024,
  \aap, 690, A206

\bibitem[{{van der Wel} {et~al.}(2014){van der Wel}, {Franx}, {van Dokkum},
  {Skelton}, {Momcheva}, {Whitaker}, {Brammer}, {Bell}, {Rix}, {Wuyts},
  {Ferguson}, {Holden}, {Barro}, {Koekemoer}, {Chang}, {McGrath},
  {H{\"a}ussler}, {Dekel}, {Behroozi}, {Fumagalli}, {Leja}, {Lundgren},
  {Maseda}, {Nelson}, {Wake}, {Patel}, {Labb{\'e}}, {Faber}, {Grogin}, \&
  {Kocevski}}]{vanderwel2014}
{van der Wel}, A., {Franx}, M., {van Dokkum}, P.~G., {et~al.} 2014, \apj, 788,
  28

\bibitem[{Virtanen {et~al.}(2020)Virtanen, Gommers, Oliphant, Haberland, Reddy,
  Cournapeau, Burovski, Peterson, Weckesser, Bright,
  {et~al.}}]{virtanen2020scipy}
Virtanen, P., Gommers, R., Oliphant, T.~E., {et~al.} 2020, Nat. Methods, 17,
  261

\bibitem[{{White} \& {Frenk}(1991)}]{white_frenk1991}
{White}, S. D.~M. \& {Frenk}, C.~S. 1991, \apj, 379, 52

\bibitem[{{White} \& {Rees}(1978)}]{white_rees1978}
{White}, S.~D.~M. \& {Rees}, M.~J. 1978, \mnras, 183, 341

\bibitem[{{White} \& {Springel}(2000)}]{white_springel2000}
{White}, S. D.~M. \& {Springel}, V. 2000, in The First Stars, ed. A.~{Weiss},
  T.~G. {Abel}, \& V.~{Hill}, 327

\bibitem[{{Yan} {et~al.}(2020){Yan}, {Jerabkova}, \& {Kroupa}}]{yan+2020}
{Yan}, Z., {Jerabkova}, T., \& {Kroupa}, P. 2020, \aap, 637, A68

\bibitem[{Zoccali {et~al.}(2014)Zoccali, Gonzalez, Vasquez, Hill, Rejkuba,
  Valenti, Renzini, Rojas-Arriagada, Martinez-Valpuesta, Babusiaux, Brown,
  Minniti, \& McWilliam}]{zoccali+2014}
Zoccali, M., Gonzalez, O.~A., Vasquez, S., {et~al.} 2014, \aap, 562, A66

\bibitem[{Zoccali \& Valenti(2026)}]{zoccali+valenti2024}
Zoccali, M. \& Valenti, E. 2026, in Encyclopedia of Astrophysics (First
  Edition), first edition edn., ed. I.~Mandel (Oxford: Elsevier), 19--37

\bibitem[{{Zoccali} {et~al.}(2018){Zoccali}, {Valenti}, \&
  {Gonzalez}}]{zoccali+2018_weighing}
{Zoccali}, M., {Valenti}, E., \& {Gonzalez}, O.~A. 2018, \aap, 618, A147

\bibitem[{{Zoccali} {et~al.}(2017){Zoccali}, {Vasquez}, {Gonzalez}, {Valenti},
  {Rojas-Arriagada}, {Minniti}, {Rejkuba}, {Minniti}, {McWilliam}, {Babusiaux},
  {Hill}, \& {Renzini}}]{zoccali+2017}
{Zoccali}, M., {Vasquez}, S., {Gonzalez}, O.~A., {et~al.} 2017, \aap, 599, A12

\end{thebibliography}

\begin{appendix}
\section{Resolution assessment}
\label{sec:appendix_reseff}
The \textsc{CIELO} sample analyzed in this article is composed of 54 galaxies, of which 19 belong to the P7 set, precisely those simulated at high resolution (L12). These were also run at an intermediate-resolution level (L11), with 19 additional counterparts not included in the main sample. Here we use P7 galaxies at L11 and L12 to assess resolution effects on knot identification.

Of the 28 knots characterized in this work, nine were identified in the P7 galaxies in the high-resolution runs (hereafter L12-knots), i.e., nine of the 19 L12-candidates. On the other hand, knots were independently classified in the P7 galaxies from the intermediate-resolution runs following the same methodology described in Section~\ref{sec:knotsincielo}, yielding four knots (hereafter L11-knots) out of the 19 L11-candidates, only two of which are in common with the nine L12-knots. However, five L11-candidates that are counterparts of L12-knots were discarded as knots due to having fewer than 500 particles (see below). This already leads us to the conclusion that the number of knots we identify is a lower limit of those that we might have identified if we had all higher-resolution simulations. Furthermore, intermediate resolution preferentially affects the identification of knots in low-mass galaxies ($\lesssim10^{9}\,\mathrm{M}_\odot$), where only the most massive systems ($\gtrsim10^{9}\,\mathrm{M}_\odot$) tend to host an identifiable knot.

\begin{figure}[h!]
    \centering
    \includegraphics[width=.9\linewidth]{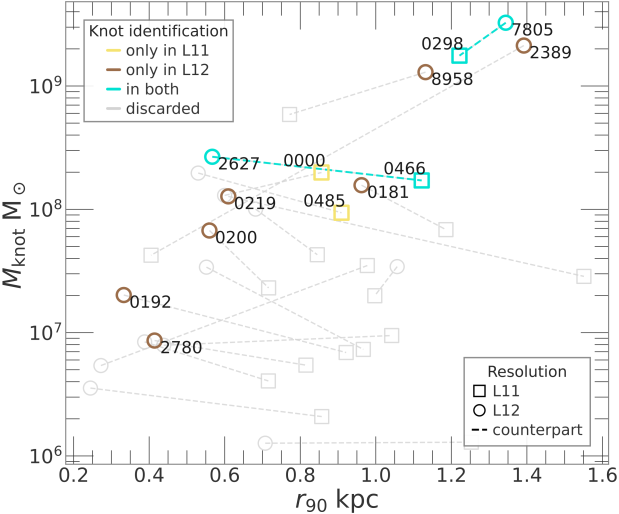}
    \caption{Total stellar mass vs. radius enclosing $90\%$ of the stellar mass for knots identified in P7 galaxies at two resolution levels (L12, circles; L11, squares). Colors indicate cross-resolution matching status (see text).}
    \label{fig:res_counterpart}
\end{figure}

Figure~\ref{fig:res_counterpart} displays the total stellar mass of L12- and L11-knots as a function of the radius enclosing $90\%$ of their stellar mass. The nine L12-knots from P7 galaxies are shown as circles and the four L11-knots as squares, and discarded candidates of both resolutions are shown in gray for completeness. Of the four L11-knots, only two match L12-knots (cyan squares and cyan circles denote the corresponding counterparts), while the remaining two (yellow squares) are unmatched L11-knots, as their host galaxies did not yield a classified knot at high resolution. The seven brown circles represent L12-knots whose host galaxies did not yield a classified knot at the intermediate resolution. As a main result, the intermediate-resolution run yields the fewest identified knots, with only four out of 19 candidates passing the selection criteria, compared to nine out of 19 in the high-resolution run. In what follows, we comment on each case.

Regarding the brown circles, five L11-candidates (in the counterparts of galaxies 0192, 0200, 0219, 2780 and 2389) were discarded for having fewer than 500 particles; one L11-candidate (counterpart of 0181) was discarded based on the concentration criteria; and one L11-candidate (counterpart of 8958) was discarded based on morphological criteria, as its mass distribution was asymmetric and off-center rather than centrally concentrated.

In one of the cyan pairs, the 7805-0298 pair, the L12-knot is slightly more extended and massive than its L11-knot counterpart. In contrast, in the other cyan pair, 2627-0466, there is a dramatic offset in $r_{90}$, where the L12-knot appears to be significantly more concentrated and slightly more massive than its L11-knot counterpart. We note that P7-2627 is precisely the galaxy where a bar and a knot coexist in L12, and interestingly, the cases marked with yellow squares correspond to galaxies hosting bars in L12. As discussed in the main text, the effect of bars on knot identification and the interplay between both structures will be addressed in Acosta-Tripailao et al. (in prep.).

Our conclusions are that the knots identified in simulated galaxies with intermediate resolution represent lower limits to the frequency of knots. In L12, our results suggest that 50\% of galaxies within the stellar mass range $10^{8.0}$--$10^{10.7}\,\mathrm{M}_\odot$ have a very well-defined, central knot. However, these numbers should be confirmed with a larger sample. In light of the comparison presented above, we acknowledge that the present simulations do not yet demonstrate numerical convergence for the knot population, and that this sensitivity is expected to propagate to the physical properties characterized throughout this work.

\section{$\left[\mathrm{O}/\mathrm{Fe}\right]$–$\left[\mathrm{Fe}/\mathrm{H}\right]$ planes}
\label{sec:appendix_alphaplane}
\begin{figure*}[ht!]
        \centering
        \includegraphics[width=\linewidth]{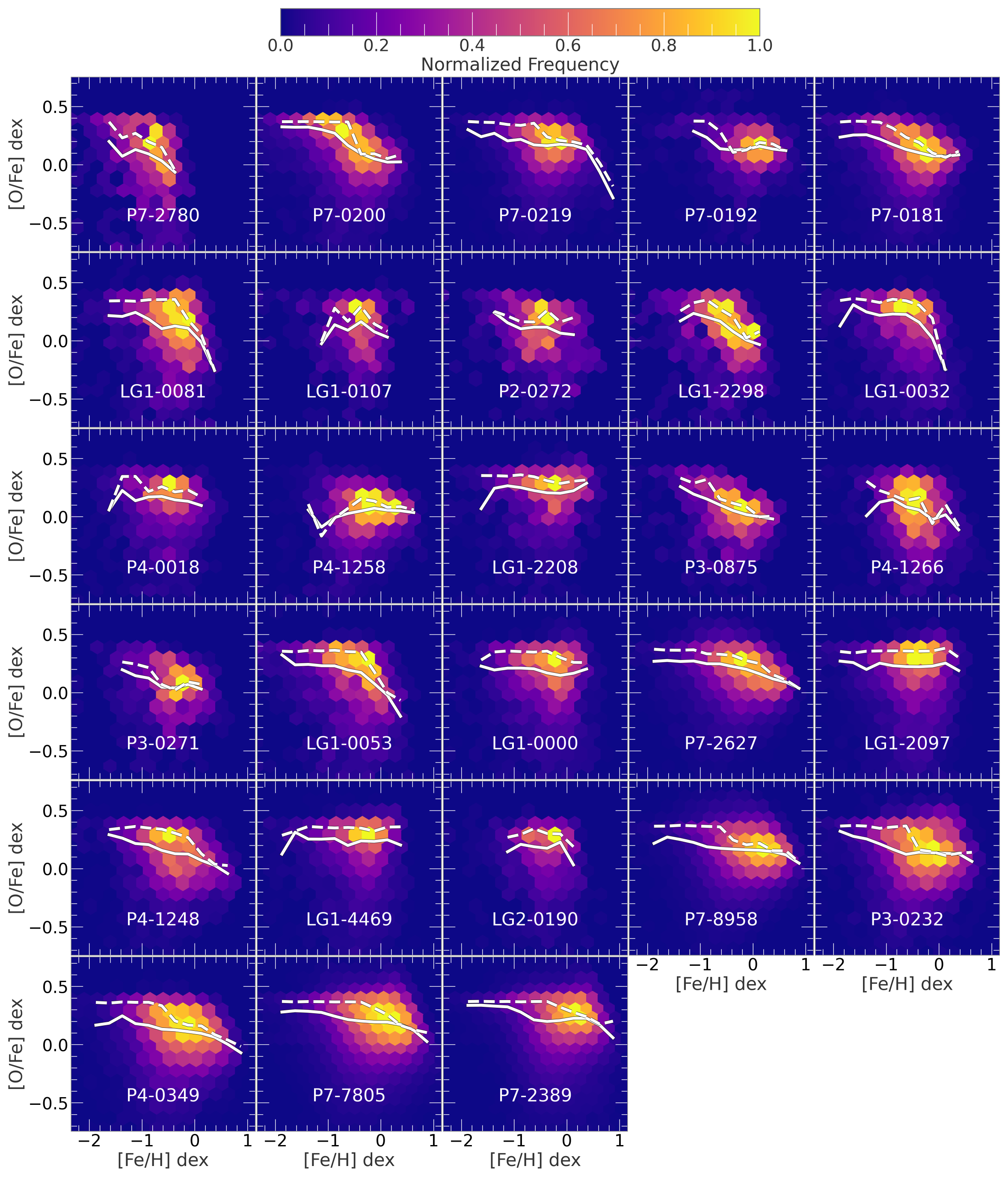}
        \caption{Individual $\left[\mathrm{O}/\mathrm{Fe}\right]$–$\left[\mathrm{Fe}/\mathrm{H}\right]$ planes of the knot stellar populations analyzed in this work. Panels are ordered by increasing host galaxy mass, with the top left being the least massive and the bottom right the most massive. Each panel shows hexbins of the $\left[\mathrm{O}/\mathrm{Fe}\right]$–$\left[\mathrm{Fe}/\mathrm{H}\right]$ plane, colored by the normalized stellar mass density, with overlaid mode (dashed) and median (solid) relations, both computed as mass-weighted statistics. These relations are derived in fixed $\left[\mathrm{Fe}/\mathrm{H}\right]$ bins using bootstrap resampling.}
        \label{fig:individual_alphaplanes}%
    \end{figure*}

We estimated the alpha-element abundance plane of knots as way of quantifying their chemical characteristics. Here, we describe the procedure and show the individual planes for the simulated knots. In Fig.~\ref{fig:individual_alphaplanes}, each panel displays hexbins of the $\left[\mathrm{O}/\mathrm{Fe}\right]$–$\left[\mathrm{Fe}/\mathrm{H}\right]$ plane, colored by the stellar mass density within each bin, normalized to the maximum bin value, together with two overlaid curves representing the mode (dashed) and median (solid) of the mass-weighted distributions~\citep{Edgeworth1888, Cormen2001} of $\left[\mathrm{O}/\mathrm{Fe}\right]$ ratios. These curves are independently derived from bootstrap resampling based on $1000$ realizations, in fixed $\left[\mathrm{Fe}/\mathrm{H}\right]$ bins. The mode closely traces the regions of highest stellar mass density in this plane, making it sensitive to the well-known ``knee'' feature clearly visible in several knot populations. However, since our goal is to follow the bulk chemical evolution of the stellar population as a whole, rather than the density peak, we adopt the median, which better captures the central tendency of the chemical abundance distribution across the full population. This choice also ensures consistency with previous works that employ the median as the reference statistic for chemical abundance relations in numerical simulations~\citep{gonzalez-jara+2025, munoz-escobar2026, iza+2025}.

\begin{figure*}
\centering
    \includegraphics[width=.8\linewidth]{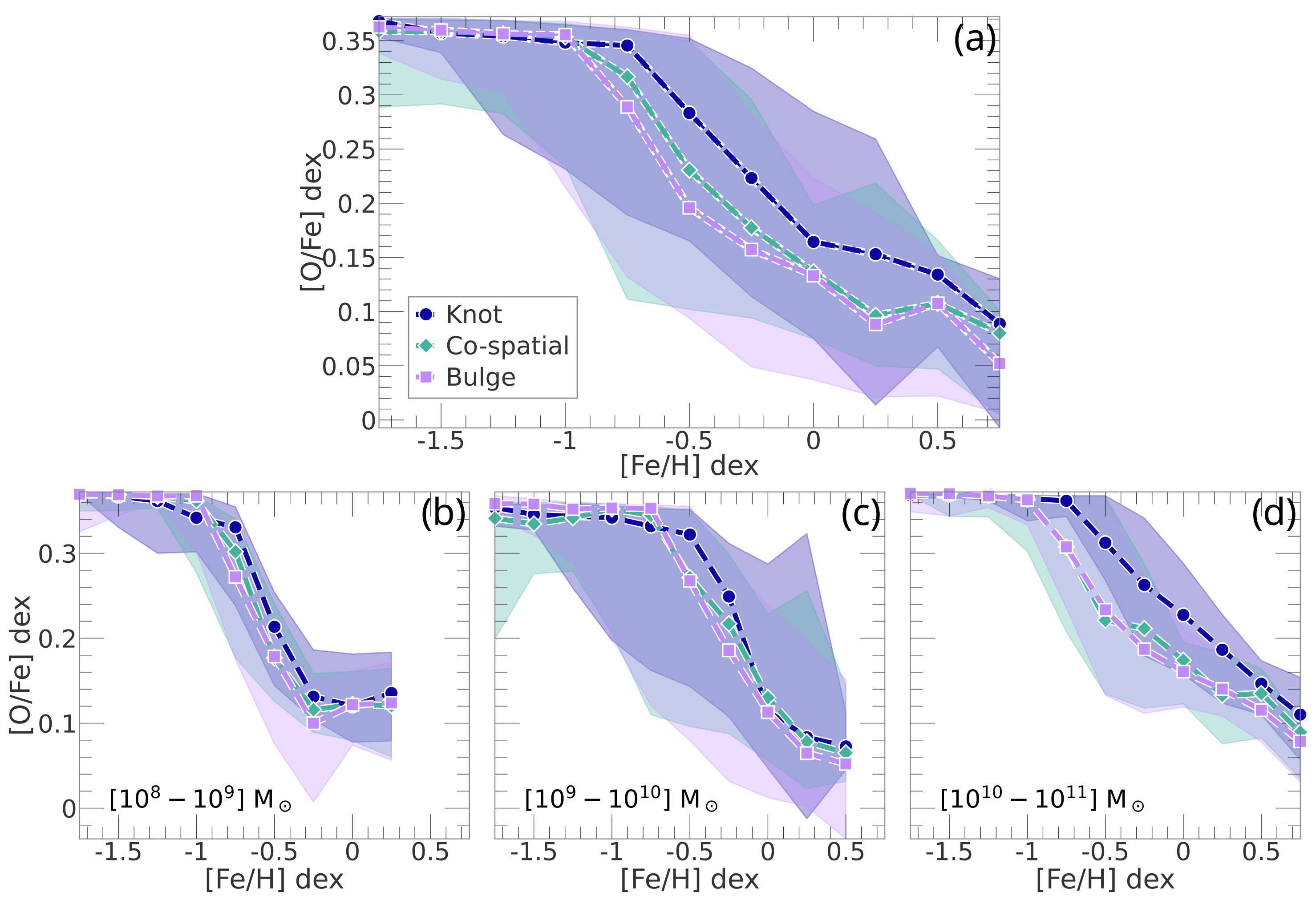}
   \caption{Panel \textbf{(a)}: Stacked $\left[\mathrm{O}/\mathrm{Fe}\right]$–$\left[\mathrm{Fe}/\mathrm{H}\right]$ relations based on the mass-weighted mode across galaxies. Dashed curves for the knot (blue), co-spatial (green), and bulge (purple) stellar populations. Shaded regions show the 16th–84th percentiles. Panels \textbf{(b--d)}: Same as panel (a), but grouped by stellar mass of the corresponding host galaxies, as indicated by the inset labels.}
    \label{fig:alphaplane_mode}
\end{figure*}

Therefore, the stacked relations shown in Fig.~\ref{fig:alphaplane_median} and Fig.~\ref{fig:alphaplane_mode} for the knots are obtained by combining the individual median (solid lines) and mode (dashed lines) relations shown in Fig.~\ref{fig:individual_alphaplanes}, respectively. The stacked relations for the co-spatial components and bulges are obtained following the same procedure. The shift traced by the mode appears primarily at higher iron abundances, in contrast to the median, which shows a shift over the full $\left[\mathrm{Fe}/\mathrm{H}\right]$ range. Although the difference traced by the mode is weaker overall, it becomes more pronounced at higher galaxy masses, as shown in Fig.~\ref{fig:alphaplane_mode}b--d.

\onecolumn

\section{Boxplot values}
\label{sec:appendix_boxplotsvalues}
The exact values displayed in the boxplots throughout this work are reported in the following tables.

\begin{table*}[ht!]
\caption{Summary statistics of $T_{50}$, $T_{90}$, and $(T_{50}-T_{90})$ for the different stellar components, shown in Fig.~\ref{fig:t5090}.}
\label{table:t5090_stats}
\centering
\begin{tabular}{l l c c c c c c}
\hline\hline
& Component & Mean & Median & Min & 25\% & 75\% & Max\\
 &  & Gyr & Gyr & Gyr & Gyr & Gyr & Gyr\\
\hline
$T_{50}$
 & Knot       & 10.71 & 11.40 & 6.56 & 9.71 & 11.60 & 12.36 \\
 & Co-spatial & 10.49 & 11.06 & 7.51 & 9.44 & 11.52 & 12.22 \\
 & Bulge      & 10.41 & 10.82 & 7.33 & 9.46 & 11.55 & 12.17 \\
\hline
$T_{90}$
 & Knot       & 7.62 & 8.06 & 1.19 & 5.52 & 10.04 & 11.91 \\
 & Co-spatial & 6.10 & 5.68 & 0.74 & 3.67 & 7.80 & 11.78 \\
 & Bulge      & 5.82 & 5.73 & 0.37 & 3.21 & 7.75 & 11.68 \\
\hline
$(T_{50}-T_{90})$
 & Knot       & 3.08 & 2.44 & 0.24 & 1.28 & 4.98 & 8.41 \\
 & Co-spatial & 4.39 & 4.36 & 0.38 & 2.86 & 6.04 & 8.87 \\
 & Bulge      & 4.59 & 4.92 & 0.46 & 2.87 & 6.05 & 8.65 \\
\hline
\end{tabular}
\end{table*}

\begin{table*}[ht!]
\caption{Summary statistics of the accreted and disk-born mass fractions for the different stellar components, shown in Fig.~\ref{fig:acc_disk_boxplots}.}
\label{table:acc_disk_fraction_stats}
\centering
\begin{tabular}{l l c c c c c c}
\hline\hline
 & Component & Mean & Median & Min & 25\% & 75\% & Max \\
\hline
Accreted mass fraction
 & Knot       & 0.14 & 0.05 & 0.01 & 0.02 & 0.16 & 0.55 \\
 & Co-spatial & 0.17 & 0.11 & 0.02 & 0.04 & 0.24 & 0.59 \\
 & Bulge      & 0.19 & 0.12 & 0.02 & 0.05 & 0.32 & 0.47 \\
\hline
Disk-born mass fraction
 & Knot       & 0.02 & 0.00 & 0.00 & 0.00 & 0.01 & 0.16 \\
 & Co-spatial & 0.06 & 0.07 & 0.00 & 0.01 & 0.08 & 0.25 \\
 & Bulge      & 0.11 & 0.10 & 0.00 & 0.05 & 0.13 & 0.41 \\
\hline
\end{tabular}
\end{table*}

\begin{table*}[ht!]
\caption{Summary statistics of the stellar mass fractions associated with different gas-origin channels, shown in Fig.~\ref{fig:gas_origin_boxplots}.}
\label{table:gas_origin_stellar_fractions}
\centering
\begin{tabular}{l l c c c c c c}
\hline\hline
Channel & Component & Mean & Median & Min & 25\% & 75\% & Max\\
\hline
Fossil bulge
 & Knot       & 0.21 & 0.21 & 0.06 & 0.13 & 0.26 & 0.37 \\
 & Co-spatial & 0.17 & 0.17 & 0.06 & 0.11 & 0.21 & 0.30 \\
 & Bulge      & 0.16 & 0.16 & 0.05 & 0.09 & 0.20 & 0.28 \\
\hline
Infall from disk
 & Knot       & 0.15 & 0.14 & 0.03 & 0.07 & 0.16 & 0.45 \\
 & Co-spatial & 0.16 & 0.14 & 0.04 & 0.10 & 0.20 & 0.45 \\
 & Bulge      & 0.18 & 0.15 & 0.04 & 0.11 & 0.23 & 0.45 \\
\hline
Infall from out
 & Knot       & 0.64 & 0.65 & 0.41 & 0.59 & 0.71 & 0.77 \\
 & Co-spatial & 0.66 & 0.68 & 0.44 & 0.61 & 0.73 & 0.77 \\
 & Bulge      & 0.67 & 0.69 & 0.43 & 0.63 & 0.72 & 0.79 \\
\hline
\end{tabular}
\end{table*}

\begin{table*}[ht!]
\caption{Summary statistics of median $\left[\mathrm{Fe}/\mathrm{H}\right]$ for stellar populations from different origin channels, shown in Fig.~\ref{fig:stellar_gaschannel_FeHOFe_boxplots}a--c.}
\label{table:star_feh_stats}
\centering
\begin{tabular}{l l c c c c c c}
\hline\hline
Channel & Component & Mean & Median & Min & 25\% & 75\% & Max\\
 &  & dex & dex & dex & dex & dex & dex\\
\hline
Fossil bulge
 & Knot       & -0.090 & -0.140 & -0.710 & -0.235 & 0.161 & 0.706 \\
 & Co-spatial & -0.170 & -0.163 & -0.728 & -0.357 & 0.093 & 0.391 \\
 & Bulge      & -0.222 & -0.256 & -1.106 & -0.411 & 0.020 & 0.314 \\
\hline
Infall from disk
 & Knot       & -0.334 & -0.340 & -0.821 & -0.454 & -0.225 & 0.339 \\
 & Co-spatial & -0.431 & -0.425 & -0.879 & -0.598 & -0.305 & 0.173 \\
 & Bulge      & -0.472 & -0.532 & -0.926 & -0.615 & -0.362 & 0.173 \\
\hline
Infall from out
 & Knot       & -0.442 & -0.450 & -0.885 & -0.565 & -0.278 & -0.031 \\
 & Co-spatial & -0.563 & -0.538 & -0.972 & -0.711 & -0.444 & -0.030 \\
 & Bulge      & -0.614 & -0.597 & -1.024 & -0.762 & -0.514 & -0.030 \\
\hline
\end{tabular}
\end{table*}

\begin{table*}[ht!]
\caption{Summary statistics of median $\left[\mathrm{O}/\mathrm{Fe}\right]$ for stellar populations from different origin channels, shown in Fig.~\ref{fig:stellar_gaschannel_FeHOFe_boxplots}d--f.}
\label{table:star_ofe_stats}
\centering
\begin{tabular}{l l c c c c c c}
\hline\hline
Channel & Component & Mean & Median & Min & 25\% & 75\% & Max\\
 &  & dex & dex & dex & dex & dex & dex\\
\hline
Fossil bulge
 & Knot       & 0.162 & 0.163 & -0.087 & 0.111 & 0.215 & 0.354 \\
 & Co-spatial & 0.154 & 0.144 & -0.028 & 0.096 & 0.183 & 0.343 \\
 & Bulge      & 0.146 & 0.138 & -0.028 & 0.081 & 0.171 & 0.337 \\
\hline
Infall from disk
 & Knot       & 0.079 & 0.087 & -0.208 & 0.022 & 0.126 & 0.289 \\
 & Co-spatial & 0.055 & 0.050 & -0.121 & -0.005 & 0.116 & 0.251 \\
 & Bulge      & 0.047 & 0.039 & -0.116 & 0.001 & 0.101 & 0.244 \\
\hline
Infall from out
 & Knot       & 0.113 & 0.117 & -0.128 & 0.049 & 0.157 & 0.315 \\
 & Co-spatial & 0.091 & 0.084 & -0.146 & 0.052 & 0.132 & 0.282 \\
 & Bulge      & 0.083 & 0.074 & -0.124 & 0.043 & 0.130 & 0.284 \\
\hline
\end{tabular}
\end{table*}

\begin{figure*}[ht!]
    \centering
    \includegraphics[width=.65\linewidth]{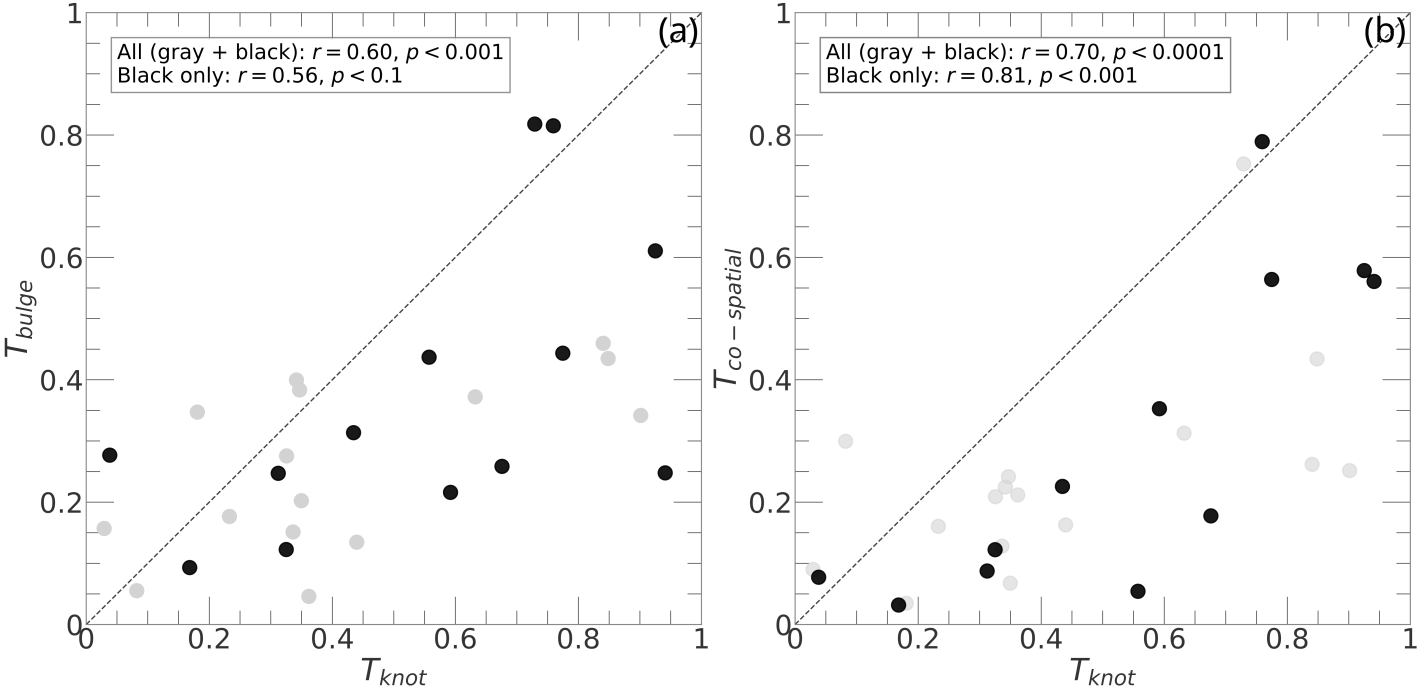}
    \caption{Triaxiality values comparing \textbf{(a)} knot vs. bulge and \textbf{(b)} knot vs. co-spatial. The dashed line shows the 1:1 relationship. All circles represent the full sample; gray-filled circles mark values falling within the shaded region of Fig.~\ref{fig:triax} ($s, q \approx 1$), while black-filled circles are the remaining subset. The Pearson correlation coefficient $r$ and its $p$-value are quoted in each panel.}
    \label{fig:triaxi11}
\end{figure*}

\FloatBarrier 
\twocolumn
\section{Triaxiality}
\label{sec:appendix_triaxiality}
Figure~\ref{fig:triax} shows contour plot of the triaxiality parameter $T$ as a function of $q$ and $s$, defining a different morphological regime: prolate, oblate, triaxial and spherical. Table~\ref{table:triaxi_regimes} shows the number of systems (knot, co-spatial, bulge) that fall into each regime. Also, Fig.~\ref{fig:triax} shows that the distributions of $s$ and $q$ differ between components indicating that some knots do not inherit the morphology of their host bulge, nor do they share it with their surroundings (co-spatial stellar particles). To verify this point, Fig.~\ref{fig:triaxi11} shows one-to-one comparisons between knots and corresponding bulges, and between knots and corresponding co-spatial components. Systems falling within the shaded region of Fig.~\ref{fig:triax} ($s, q \approx 1$) are marked in gray-filled circles. Both comparisons point to positive correlations for the full sample ($r = 0.60$ and $r = 0.70$, with $p$-values $<0.001$ and $<0.0001$, respectively), yet the considerable scatter away from the 1:1 relationship indicates that a fraction of knots are structures largely independent of both their host bulge and co-spatial stellar environment.

\begin{table}[ht!]
\caption{Number and percentage of systems in each shape regime for the different stellar components, shown in Fig.~\ref{fig:triax}.}
\label{table:triaxi_regimes}
\centering
\begin{tabular}{l c c c}
\hline\hline
Regime & Knot & Co-spatial & Bulge\\
\hline
Prolate   & 6  (21.4\%) & 1  (3.6\%)  & 2  (7.1\%)  \\
Triaxial  & 3  (10.7\%) & 4  (14.3\%) & 9  (32.1\%) \\
Oblate    & 4  (14.3\%) & 15 (53.6\%) & 14 (50.0\%) \\
Spherical & 15 (53.6\%) & 8  (28.6\%) & 3  (10.7\%) \\
\hline
\end{tabular}
\end{table}

\end{appendix}
\end{document}